\documentclass[useAMS,usenatbib,usegraphicx]{mn2e}
\usepackage{color}
\usepackage{colortbl}
\usepackage{journals}

\begin{document}
\def\mpc{h^{-1} {\rm{Mpc}}} 
\def\up{h^{-3} {\rm{Mpc^3}}} 
\def\uk{h {\rm{Mpc^{-1}}}}
\def\kms {\rm{km~s^{-1}}} 
\def\apj {ApJ} 
\def\aj {AJ} 
\def\mnras {MNRAS} 
\def\hii {\mbox{H\,{\sc ii}} }

\title[Compact Groups from the Millennium Simulations]{Compact groups from
  the Millennium Simulations: I. Their Nature and the completeness of the
  Hickson sample}
\author[D\'{\i}az-Gim\'enez \& Mamon] {Eugenia D\'{\i}az-Gim\'enez$^{1,2}$ \& 
Gary A. Mamon$^{3,4}$ \\ 
$1$ Instituto de Astronom\'{\i}a Te\'orica y Experimental, IATE, Observatorio
Astron\'omico, Laprida 854, C\'ordoba, Argentina\\ $2$ Consejo de
Investigaciones Cient\'{\i}ficas y T\'ecnicas de la Rep\'ublica
Argentina (CONICET) \\ $3$ Institut d'Astrophysique de Paris (UMR 7095: CNRS
\& Univ. Pierre \& Marie Curie) 98 bis Bd. Arago, F--75014 Paris, France
\\ $4$ Astrophysics \& BIPAC, Department of Physics, University of Oxford, Oxford
OX1 3RH, UK} 

\date{\today} \maketitle 

\begin{abstract}

We identify compact groups of galaxies (CGs) within mock galaxy catalogues from 
the Millennium Simulation at z=0 with the semi-analytic models (SAMs) 
of galaxy formation 
of  Bower et al., Croton et al. and De Lucia \& Blaizot.
CGs are identified using the same 2D criteria as those visually applied by 
Hickson (1982) to his CGs (HCGs), but with a brightest galaxy magnitude
limit, and we also add the important effect of
observers blending close projected pairs. Half of the mock CGs identified in projection 
contain at least 4 accordant velocities (\emph{mvCG}s), versus
70\% for HCGs.
In comparison to \emph{mvCG}s, the HCGs are only 8\% complete at distances $< 9000
\, \rm km \, s^{-1}$, 
missing the CGs with small angular sizes, a strongly dominant
galaxy, and (for the second SAM) the \emph{mvCG}s that are fainter and those
with lower surface brightness.
10\% of the mock \emph{mvCG}s are identical to the parent virialized group, meaning
that they are isolated, while the remainder are embedded in their parent
virialized groups.
We explore different ways to determine the fraction of physically dense
groups given the data from the simulations.
Binding energy criteria turn out to be inapplicable given the segregation between
galaxies and dark matter particles. We rely instead on the combination of the
three-dimensional length of the CGs (maximum real space galaxy separation) 
and their elongation along the line-of-sight (ratio of maximum line-of-sight
to maximum projected separations), restricting ourselves in both
cases to smallest quartets within the CGs.
We find that 
between 64\% and 80\%
(depending on the SAM) of
the \emph{mvCG}s have 3D lengths shorter than $200 \, h^{-1} \, \rm kpc$, between
71\% and 80\% have line-of-sight elongations less than 2, while between 59\%
and 76\% have either 3D lengths shorter than $100 \, h^{-1} \, \rm kpc$ or
both lengths shorter than $200 \, h^{-1} \, \rm kpc$ and elongations smaller
than 2.
Therefore,
chance alignments (CAs) of galaxies concern at most 40\% of the \emph{mvCG}s.
These CAs are mostly produced 
from larger host groups, but a few have galaxies extending a few Mpc beyond the 
host group. 
The \emph{mvCG}s built with the Hickson selection (respectively without the
close projected
pair blending criterion) have 10\% higher (lower) fractions of
physically dense systems. 

\end{abstract} 
\begin{keywords} galaxies: clusters: general -- methods: data analysis 
\end{keywords}
\section{Introduction} 
Compact Groups (CGs) are small, relatively isolated systems of typically 
four or five luminous galaxies in close proximity to one another. The first example 
of a CG was found by \cite{Stephan1877}. Several catalogues of CGs are now 
available: \cite{Rose77} and \cite{Hickson82}
visually identified CGs on POSS~I photographic plates.
After the Hickson compact group (HCG) catalogue, 
several CG catalogues have been automatically extracted
from galaxy catalogues, themselves automatically extracted from photographic
plates: from the
COSMOS/UKST Southern Galaxy Catalog \citep*{PIM94,Iovino02}, from the
DPOSS catalogue
\citep{Iovino03,decarvalho05} or CCD frames from the Sloan Digital Sky Survey
(SDSS) photometric
catalogue \citep{Lee+04}.
CG catalogues have also been extracted from galaxy catalogues in redshift
space:
 from the CfA2 
\citep{Barton+96},
Las Campanas \citep{AT00}, and SDSS \citep{Deng+07,McCPES09} surveys, as well as from
the 3D UZC Galaxy Catalog \citep{FK02}.
CGs are so compact that the median projected galaxy separation in HCGs is
only $39 \, h^{-1} 
\, \rm kpc$ \citep{HMdOHP92}.

HCGs have been studied in detail, in particular their internal structures,
shapes, morphologies, luminosities, and environments
\citep{Hickson82,HNHM84,Mamon86,HKH88,HR88,MdOH91,Zepf93,Moles+94,PIM94,KF04,TPT06}.
To summarise, these studies indicate that CG galaxies have star formation properties,
colours and morphological mixes that lie in between binary galaxies and
isolated ones.

The nature of the CGs has been a puzzling matter for quite some time. 
How can a few bright galaxies coexist within less than 100 kpc,
given that galaxies are expected to merge fast in such systems
\citep*{CCS81,Barnes85,Mamon87,BCL93}?
 There are
three schools of thought on this matter. One view is that compact groups are 
recently formed dense systems that are about to coalesce into a single galaxy
\citep{HR88}. 
The galaxies lost in the merger may be replenished by
galaxies in the loose group environment \citep*{DGR94}, and the predicted 
rate of formation of CGs appears to be sufficient to explain the observed
frequency of HCGs \citep{Mamon00_IAU174}.
The 
second view states that CGs may be transient unbound cores of looser groups 
\citep{Rose77,RDGH94,TYT01}.
And the third scenario places CGs as chance alignments of galaxies 
along the line of sight 
within larger loose groups (\citealp{Rose77} for CGs elongated in
projection; \citealp{Mamon86} and \citealp{WM89} in general), clusters
\citep{WM89} and  
cosmological filaments \citep*{HKW95}.
In this scenario,
the numerous signs of interaction and star formation is explained by the
frequent occurrence of 
binaries and triplets in the chance alignments \citep{Mamon92_DAEC}.

If CGs are physically dense, their dynamical times should be short (1\% of
the age of the Universe), and the hot intra-group gas should trace a smooth
gravitational potential.
The launch of X-ray observatories with good 
sensitivity in the soft X-ray band (ROSAT, ASCA, Chandra and XMM-Newton)  
has led to the detection of hot X-ray emitting gas from 
many CGs. Since the X-ray emissivity scales as the square of the gas density,
X-ray emission is less prone to projection effects than optical surveys (but
see \citealp*{OLH95}). However, although 22 HCGs were detected out of 32
pointed observations \citep{PBEB96}, it is not clear what the global fraction
of detections would be on the full sample of HCGs (69 [92] groups with at
least 4 [3] accordant velocities, according to \citealp{HMdOHP92}). Moreover,
some of the detected groups appear clumpy (e.g. HCG~16 according to 
\citealp{DSM99}), which strongly suggests their unvirialized state.
 
The distinction between compact groups that are dense in 3D, or chance
alignments within loose groups or longer filaments is difficult, because
redshift space distortion introduces uncertainties
in the computation of the line of sight coordinate which might result in 
misidentified compact configurations.
For a group with line of sight velocity dispersion $\sigma_v$, the redshift
  distortion will amount to a spread of $\delta r_z = \sigma_v/H_0$ in the line
  of sight 
  coordinate. Assuming that the square velocity dispersion is half the square
  circular velocity at the virial radius, $\sigma_v^2 = (1/2)\,GM(r_v)/r_v$
  (appropriate for an $\rho \propto 1/r^2$ density profile), one finds
$\delta r_z/r_v = (1/2)\,\sqrt{100} = 5$ if the virial radius is defined where 
the mean density at that radius is 100
  times the critical density of the Universe.
So redshift space distortions prevent measuring distances within virial
  systems (see also the Introduction of \citealp{WM89}).

Nevertheless, there is one CG meeting the HCG criteria discovered by one of us
\citep{Mamon89} that is so close (within the Virgo cluster) that
surface brightness fluctuation distance measurements by \cite{Mei+07} are
able to settle the issue of its nature: \cite{Mamon08} concludes that this
CG is a chance alignment of galaxies along the line of sight, at least 440 kpc
and most probably 2 Mpc long. 

In summary, even though many efforts have been devoted to look for an 
explanation about the nature of CGs, the debate is still wide open.

The advent of increasingly realistic cosmological simulations now allow one
to distinguish whether CGs
are truly dense in 3D, or caused by chance alignments within looser groups,
filamentary structures, or the general field.
In an early pioneering attempt, \cite*{HKW95}, who identified galaxies as
dense knots of cold gas in their $N$-body + 
hydrodynamical simulation, and searched for CGs in
redshift space in many viewing directions.
They found four CGs with at least
4 accordant velocities, all of which
were longer than $2 \, h^{-1} \, \rm Mpc$ along the line of sight (one
was as long as $4 \, h^{-1} \, \rm Mpc$), and yet presented accordant
velocities, despite the (Hubble law) 
stretching of velocities caused by their elongation
along the line of sight.
The analysis of \citeauthor{HKW95} suffers from several drawbacks (according
to present-day standards for cosmological simulations):
the simulation box was small ($44 \, h^{-1} \, \rm Mpc$
wide), the mass resolution was poor (their simulation had $32^3$ dark
matter particles and $32^3$ gas particles, and their galaxies were identified
with as few as 8 gas particles), and the spatial resolution was poor
(the dark matter particles had a softening length of $10 \, h^{-1} \, \rm
kpc$). Furthermore, the identification of galaxies with knots of dense gas was
not optimal, especially that feedback from supernovae and active galactic
nuclei were not incorporated.

In this work, we  quantify the fraction of CGs that can be considered
as physically dense entities in samples of automatically identified CGs,
based upon more realistic cosmological $N$ body simulations.
At present, one can build realistic CGs
in two ways: 
1) from dissipationless cosmological simulations on top of which galaxies
are painted using fairly complex semi-analytical galaxy formation/evolution
models;
2) from hydrodynamical codes that resolve galaxies.
We have chosen the first approach and use for this purpose the largest
cosmological $N$ body simulation ever performed (in 2006, 
when the present study began), 
the Millennium Run
\citep{Springel+05}, on which
galaxies were identified in three ways, using three different 
state-of-the-art semi-analytic models
(SAMs) of
galaxy formation
\citep{Bower+06,Croton+06,dLB07}. 

These three galaxy samples provide an opportunity to test both, 
projection effects and the real nature of systems 
identified using standard algorithms like that proposed by \cite{Hickson82}.
The CGs are identified in mock redshift-space
catalogues constructed from the real-space galaxy sample
derived with the semi-analytical model, from the Millennium Run.

In comparison with the analysis of \citeauthor{HKW95}, our study is based upon
a simulation in a box whose volume is over 1 million times greater, with
30
thousand times as many particles, 25 times finer  mass
resolution and a softening scale 4 times smaller. However, the
simulation we use does not contain gas particles, so the galaxy parameters
are highly dependent on the physics of galaxy
formation and evolution of the three SAMs that we analyse.

We focus here on the HCG catalogue, 
which is by far the best studied sample of Compact
Groups. 

The layout of this paper is as follows. In Section~\ref{sample}, we present the
different steps for the construction of the mock CG
catalogue and Sect.~\ref{classes} describes how the resulting CGs are classified.
The 
conclusions are summarised and discussed in Section~\ref{discus}.
Once our analysis was well advanced, we learnt about the work of
\cite{McCEP08}, who performed a similar analysis of the properties of
Hickson-like CGs from the \cite{dLB07} galaxy catalogue, and found that 70\% of
the mock CGs selected in projection were caused by chance alignments of galaxies.
We highlight in Sect.~\ref{compare} the similarities and several 
important differences between our two studies.

\section{Construction and classification of the compact group sample}
\label{sample}

\subsection{Observed compact group sample}
\label{hcgsample}
We use the HCG catalogue of compact groups, with
photometry measured by \cite*{HKA89} in the $B$ and (presumably Johnson) $R$
bands. 
\cite{Hickson82} found 100 HCGs, and
\cite{HMdOHP92}, who measured the redshifts for virtually all galaxies,
built a velocity sample of \emph{vHCG}s 
by eliminating galaxies lying at more
than $1000 \, \rm km \, s^{-1}$ from the group's median velocity.
In this manner, 
they obtained 92 HCGs with at least 3
accordant velocities and 69 HCGs with at least 4 accordant velocities.
We extracted the photometry and velocities using
Table {\tt VII/213/galaxies} in
VizieR\footnote{http://webviz.u-strasbg.fr/viz-bin/VizieR} \citep*{OBM00}.
This database contains the velocities for all galaxies except 6. 
We found the redshifts for these  6 galaxies 
(Table~\ref{nedvs}) in the NASA/IPAC Extragalactic Database
(NED)\footnote{http://nedwww.ipac.caltech.edu}.

\begin{table}
\centering
\caption{Additional HCG galaxy redshifts found in NED\label{nedvs}}
\tabcolsep 1pt
\begin{tabular}{cccl}
\hline
Galaxy & $v$ & $\epsilon(v)$ & Reference \\
&  $\left (\rm km \, s^{-1}\right )$ & $\left (\rm km \, s^{-1}\right )$ & \\
\hline
\ \,18b &  \ \,4105 &\ \,25 & \cite{Falco+99}\\
\ \,19c &  \ \,4253 &\ \,23 & \cite{dCRCZ97} \\
\ \,19d & 20443 & \ \,26 & \cite{dCRCZ97} \\
\ \,51g &  \ \,7532 &\ \,41 & \cite{RC3} \\
\ \,57h &  \ \,9240 & 105 & \cite{Hickson93,BdCG98} \\
100d &  \ \,5590 &\ \,32 & \cite{Hickson93} \\
\hline
\end{tabular}

Notes: The radial heliocentric velocities and their errors are given in columns
2 and 3, respectively. For HCG57h, the velocity is from the first reference,
while the error is from the second.
\end{table}

For future comparisons with the SDSS,
we choose a Johnson $R$-band magnitude limit 
of 17.44
that mimics the SDSS
spectroscopic magnitude limit of $r
< 17.77$ (see appendix \ref{appmags}).
We measure the total raw magnitude $R_T$ and the extinction-corrected
magnitude $R_T^0$ using
\begin{eqnarray}
R_T &\simeq& B_T - (B-R) \ ,\nonumber \\
R_T^0&\simeq& B_T^0 - (B-R) + E_{B-R} \ , \nonumber \\
E_{B-R} &=& \left (B_T-B_T^0\right)\,\left (1-{A_R/A_V\over A_B/A_V}\right)
\ ,
\nonumber 
\end{eqnarray}
where $B_T$, $B_T^0$ and $B-R$ are the raw total blue magnitude,
extinction-corrected total blue magnitude and isophotal $B-R$ colour, all
given in VizieR.

We first note that only 83 HCGs 
out of the original 
99\footnote{We have omitted group
HCG 54, which is the HCG with the smallest 
projected radius, as it appears to be either a group of \hii
regions in a single galaxy \citep{Arkhipova+81} or the end result of the
merger of two 
disk galaxies \citep{VerdesMontenegro+02}.} 
have at least 4 galaxies whose extinction-corrected $R$-band magnitudes are
within 3 mag from the brightest one.\footnote{This is also clear in the $B_T$ and
$B_T^0$ magnitudes given in \citealp{HKA89}, although this was not discussed by
these authors, but was also noted by \cite{Sulentic97}.}
Six of the HCGs do not satisfy the HCG isolation criterion \citep{Sulentic97}
and these were also omitted from our sample.
On the other hand, we re-inserted into our sample HCG~31, which has
additional members \citep{Sulentic87}, among which two
additional members within 3 magnitudes from the brightest member:
galaxies G and Q, for which we adopt the $R$-band photometry of \cite{RHF90},
while the radial velocities are taken from \cite{MendesdeOliveira+06}, and
convert to $R_T^0$ using the median difference for other galaxies: $R_T^0 = R
- 0.38$.

We are left with 72 HCGs whose brightest magnitude satisfies $R_b <
17.44-3 = 14.44$,
thus ensuring completeness out to $R=17.44$.
We call this the \emph{pHCG} sample (for HCGs defined in projected space). 

Only 52 among the 72
\emph{pHCG}s
have at least 4 galaxies within $1000 \, \rm km \, s^{-1}$ from the
median group velocity 
(hereafter the \emph{vHCG} sample, for
velocity-selected HCG).

\subsection{Basic scheme for mock compact group samples}
Our mock catalogues of CGs are built in several steps, in which we:
\begin{enumerate}
\item simulate the gravitational evolution of a large piece of the Universe,
  represented by collisionless (dark matter) particles;
\item attach galaxies to the simulation with a semi-analytical galaxy
  formation model
\item convert to a mock galaxy catalogue in redshift space;
\item convert to a mock 2D CG catalogue (hereafter \emph{mpCG} for mock CG in
  projection), by applying the HCG selection criteria; 
\item convert the \emph{mpCG} catalogue 
to a velocity-filtered mock CG catalogue (hereafter, \emph{mvCG}
  for mock velocity-filtered CG), by removing galaxies with discordant
  redshifts;
\item convert the \emph{mvCG} catalogue to a mock velocity-filtered HCG catalogue (hereafter,
  \emph{mvHCG} for mock velocity-accordant Hickson Compact Group), by randomly
  selecting groups according to the completeness of the HCG as a function of
  group surface brightness, brightest galaxy magnitude and its contribution
  to the total group
  luminosity.
\end{enumerate}
The last step is motivated by the strong incompleteness of the HCG catalogue
in surface magnitude and
galaxy magnitude (see Sect.~\ref{redshift}, below).

A list of the different acronyms used to 
refer to the different samples is provided in Table~\ref{acronyms}. 

\begin{table*}
\begin{center}
\caption{List of acronyms used throughout this work}
\label{acronyms}
\tabcolsep 3.5pt
\begin{tabular}{ll}
\hline
CG   & general compact groups\\
\hline
HCG  & Hickson compact groups\\
\emph{pHCG} & HCGs that strictly meet the Hickson (1982) criteria $+R_b\le14.44$\\
\emph{vHCG} & velocity accordant \emph{pHCG}s\\
\hline
\emph{pmpCG} & particle mock projected compact groups, which strictly meet the Hickson (1982) criteria $+R_b\le14.44$\\
\emph{mpCG} & observable mock projected compact groups (same as
\emph{pmpCG}s, but accounting for galaxy confusion)\\ 
\emph{pmvCG} & particle mock velocity accordant compact groups\\
\emph{mvCG} & observable mock velocity accordant compact groups (same as
\emph{pmvCG}s, but accounting for galaxy confusion)\\
\emph{mvHCG} & observable mock velocity accordant compact groups with Hickson's biases\\
\hline
CA  & chance alignment of galaxies\\
CALG  & chance alignment of galaxies within looser groups\\
CAF  & chance alignment of galaxies within filaments \\
\hline 
\emph{PG3D} & Parent groups identified in real space\\
\hline 
\end{tabular}
\parbox{\hsize}{ 
}
\end{center}
\end{table*}

\subsection{Dark matter particle simulation}
We use the Millennium Simulation, which is a 
cosmological Tree-Particle-Mesh (TPM, \citealp{Xu95})
$N$-body simulation \citep{Springel+05}, which
evolves 10 billion ($2160^3$) dark matter particles in a $500
\, h^{-1}\,\rm Mpc$ periodic box, using a comoving
softening length of $5 \, h^{-1} \,
\rm kpc$.\footnote{The Millennium Simulation, run by the Virgo Consortium, is
  publicly available at http://www.mpa-garching.mpg.de/millennium} 
The cosmological parameters of this simulation correspond to a flat 
cosmological model with a non-vanishing cosmological constant ($\Lambda$CDM):
$\Omega_m=0.25$, $\Omega_\Lambda=0.75$, $\sigma_8=0.9$ and $h=0.73$. 
The simulation was started at $z = 127$, with the particles initially
positioned by displacing particles initially in a glass-like distribution
according to the $\Lambda$CDM primordial density fluctuation 
power spectrum.
The $10^9$ particles of mass $8.6\times10^8 h^{-1} M_\odot$
are then advanced with the TPM code, 
using $11\,000$ internal time-steps, on a 512-processor supercomputer.
The positions and velocities of
the 10 billion particles were saved at 64 epochs (leading to nearly 20 TB
of data).

\subsection{Modelling galaxies}
We consider the $z=0$ outputs from three different SAMs of galaxy formation by 
\cite{Bower+06}, \cite{Croton+06}, and \cite{dLB07}, (B06, C06 and DLB,
respectively), where each model was applied in turn to the outputs of the
Millennium Simulation described above.
Note that, while the B06 and C06 models were developed independently, 
the DLB model is essentially the same as the C06 model, except that
the merger rate is reduced by a factor 2, the magnitudes are derived using
spectral synthesis models based upon a different initial mass function 
(\citealp{Chabrier03} instead of \citealp{Salpeter55}) with
fewer low mass stars, and the treatment of
radiative transfer to dust is much more refined.

The three SAMs produce 
galaxy 
positions, velocities, as well as absolute magnitudes (in five or more
optical and near-infrared wavebands, all including Johnson $R$),
as well as other quantities.
To summarise, the branches of the halo merger tree (produced by the
Millennium Simulation) are followed forward in
time, and the following astrophysical processes are applied: 
gas infall and cooling, early reheating of the intergalactic medium by
photoionization, star formation, black hole growth, AGN and supernova feedback,
galaxy mergers, spectro-photometric evolution, etc.
The model parameters have been adjusted to produce
a good match to the observed properties of local galaxies.
In these SAMs, AGN feedback is responsible for the absence 
of cooling flows in rich clusters, for the cut-off at the bright end of 
the galaxy luminosity function and for the number density properties of 
the most massive galaxies at all redshifts.
Also, the early reheating of the IGM by photoionization is responsible for
suppressing gas cooling in halos 
below a circular velocity that is independent of redshift (or nearly so).
The 3 SAMs produce $z=0$ galaxy luminosity functions that are in good agreement 
with observations  in both the $b_J$ and $K$ wavebands,\footnote{The
  $z=0$ luminosity function of the \cite{dLB07} model is given by
  \cite{BDLT07}.}
with an excess of galaxies at very bright luminosities for all 3 models and
a slight excess at faint luminosities for the C06 and DLB models.
Moreover, the B06 SAM provides several other observational
predictions:  
the $b_J$ and $K$ galaxy 
luminosity functions at higher redshifts, the  global history of star
formation, and the local black hole mass vs. bulge mass relation.

All three SAMs produce around 10 million galaxies at $z=0$.
The galaxy samples appear to be
complete at least to $M_{R}- 5\,\log h < -17.4$ with stellar masses
$M_*> 10^9 h^{-1}{\cal M}_\odot$ (C06) or
$M_*>3\times 10^8 h^{-1}{\cal M}_\odot$ (B06, DLB).

Each of the three SAMs has its strengths and weaknesses.
The B06 model computes galaxy mergers by inferring the
positions of galaxies in their halo through typical values of their energies
and angular momentum in units dimensioned to the virial scales of the halos.
In contrast, the C06 and DLB models have the
advantage of estimating the merger rates directly from the positions of
subhaloes in the dark matter simulation. They both use the same analytical formula for
the orbital decay time by dynamical friction once the subhalo masses fall
below their resolution limit, where the DLB time is twice the C06 time, which
itself matches almost perfectly the decay time
that \cite{Jiang+08} calibrated on high-resolution cosmological
hydrodynamical simulations.
Unfortunately, C06 do not provide the galaxy merger trees, so it is difficult 
to derive the history of star formation of a
given galaxy.
Also, while B06 find a Red Sequence with increasing red
colours for increasingly higher stellar masses, 
the C06 catalogue shows a 
colour-luminosity relation
for the Red Sequence galaxies that flattens at high luminosity, contrary to
observations, and a similar effect is seen in the DLB galaxy
output (as shown by \citealp{BDLT07}).
Still, the B06 colours are too blue and fit somewhat less
well the SDSS colour distribution than do the DLB colours
\citep{MJG08}.
The SAM of \cite{Cattaneo+06} reproduces better the colours of galaxies, but its
output is not public and the galaxy positions are determined
stochastically (like B06) rather than by following the dark matter subhaloes
(like C06 and DLB).
The DLB catalogue produces galaxies whose present-day 
small-scale segregation of recently formed stellar mass is too
large, while that of B06 matches well that
observed with the SDSS (\citeauthor{MJG08}). This is surprising given that
the B06 model treats galaxy mergers using stochastic
positions rather than the positions of the subhaloes with which the galaxies
are associated (see above). However, the DLB model predicts
a little better than B06 
the analogous segregation for intermediate age ($0.2-0.5\,\rm Gyr$)
stellar mass (\citeauthor{MJG08}).
But the present-day galaxy merger rate of DLB 
appears too low,
while that of B06 matches well the observations of the
frequency of galaxy pairs \citep{Mateus08}.

In summary, it is very difficult to decide which of the three SAMs is most
appropriate for our study of CGs, and we therefore decided to analyse the
outputs of all
three of them. 
We will find and illustrate several important differences in the properties of
mock CGs predicted from these three models.

\subsection{Mock galaxy catalogues}
\label{mockgals}

Using the snapshots at $z=0$, we construct mock catalogues in redshift space.
For the three SAMs,
we obtain redshifts by adding the Hubble flow 
to the peculiar velocities projected in the line of sight direction.
We compute the observer-frame galaxy apparent 
magnitudes from the rest-frame absolute magnitudes
provided by the semi-analytical model. These apparent 
magnitudes are converted to the observer frame
using tabulated $k+e$ corrections \citep{Poggianti97}.

Our mock catalogue is constructed by viewing the full volume of the
simulation box from one of its 8 vertexes ($z_{\rm max} \sim 0.17$, $\pi /2\,\rm
sr = 5156
\,\rm deg^2$ )
We set
an apparent magnitude limit $R=17.44$, equal to the limit we set on
the HCG groups to match the SDSS spectroscopic
catalogue for later comparisons (see Sect.~\ref{hcgsample}).

In order to increase the statistical significance of our results, we 
considered eight observers situated at the eight vertexes of the simulation box, 
and all the identification procedures were performed on these 
eight samples to finally combine the resulting CGs into one larger sample.
These eight samples are almost fully statistically independent, since
3/4 of the mock CGs selected in the cone (see Fig.~\ref{test}) lie within half
the box size ($256 \, h^{-1} \, \rm Mpc$) 
(see upper right panel of Fig.~\ref{distribs}).
Table~\ref{mocks} summarises the main properties of the three mock galaxy
catalogues seen from one of its vertexes.

The completeness of our magnitude+volume limited mock catalogues might be an important
issue that could bias the results. 
The implications on the results of using magnitude+volume limited samples
will be carefully tested in Sect.~\ref{testing}.

\begin{table}
\begin{center}
\caption{Mock galaxy catalogues ($R<17.44$)}
\label{mocks}
\tabcolsep 3.5pt
\begin{tabular}{lrccc}
\hline 
Mock & \multicolumn{1}{c}{\#} & $z_{\rm med}$ & $\overline n_{90} $  \\ 
 & &  & $ (h^3\,\rm Mpc^{-3}$) \\
\hline 
\cite{Bower+06} & $556\,224$ & 0.0998 & 0.062 \\
\cite{Croton+06} & $446\,153$ & 0.0969 &  0.059 \\
\cite{dLB07} & $1\,034\,619$ &  0.1114 & 0.089 \\
\hline 
\end{tabular}
\parbox{\hsize}{ 
Notes: \#: number of galaxies seen from a single vertex of the simulation box, $z_{\rm med}$: median redshift, $\overline n_{90}$:
space number density within $90 \, h^{-1} \, \rm Mpc$,
defined in equation~(\ref{n90}).
}
\end{center}
\end{table}

We identify regular groups of galaxies  
in the simulation box by applying a Friends-of-Friends 
(FoF) algorithm in real space \citep{DEFW85} to the galaxies.
We adopt a linking length of 
$l=0.17 \ n^{-1/3}$. 
where $n$ is the mean space density of galaxies. 
The factor 0.17 roughly corresponds to an over- 
density of 100 relative to the critical density of the Universe, 
roughly the minimum overdensity (hence maximum radius) 
where cosmological structures are in dynamical equilibrium 
(\citealp{BN98}, but recent work by \citealp{CPKM08} shows that on the mass
scales of groups, the radius of equilibrium is roughly 30\% greater).
We denote these groups the \emph{PG3D}s for \emph{Parent groups
  selected in real space}, and will later check if the mock CGs extend
beyond these \emph{PG3D}s. 


\subsection{Mock compact groups selected in projected space}
\label{criteria}

In this work, we use an automated \emph{mpCG} search algorithm 
very similar to that described by \cite{Hickson82},
applied to the three mock galaxy catalogues.
The algorithm 
defines as \emph{mpCG}s those systems that satisfy the following conditions:
\begin{enumerate}
\item $4\le N \le 10$ (population)
\item $\mu_R < 26 \rm\,mag\,arcsec^{-2}$ (compactness)
\item $\theta_{\cal N} > 3 \,\theta_{\rm G}$ (isolation)
\item $R_{\rm brightest} \le 14.44$ (flux limit) 
\end{enumerate}
where
\begin{itemize}
\item $N$ is the total number of galaxies whose $R$-band magnitude satisfies
  $R<R_{\rm brightest}+3$, where $R_{\rm brightest}$ is the magnitude of the
  brightest galaxy;
\item $\mu_R$ is the mean $R$-band surface magnitude,
 averaged over the smallest circle circumscribing the galaxy centres.
\item $\theta_{\rm G}$ is the angular diameter of this smallest circumscribed
circle;
\item $\theta_{\cal N}$ is the angular diameter of the largest 
concentric circle that contains no other galaxies within this 
magnitude range or brighter;
\end{itemize}
Note that the fourth criterion (which implies $R_{\rm faintest} \leq 17.44$)
  was not considered by \cite{Hickson82}.  
This restriction is fundamental 
for avoiding selection biases, as will
be demonstrated in Sect.~\ref{testing}.

\begin{figure}
\centering
\includegraphics[width=\hsize]{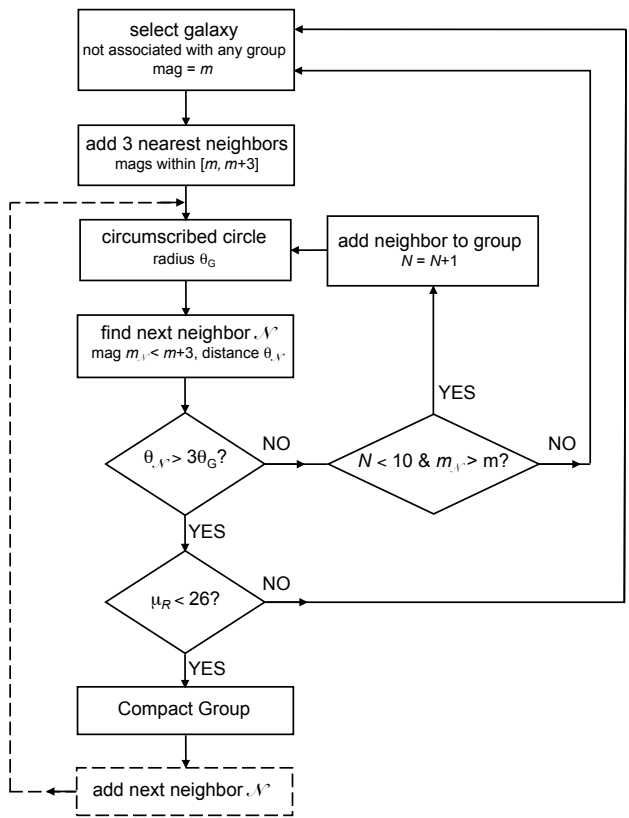}
\caption{Flowchart of the selection of projected compact groups. All
  magnitudes must be brighter than our chosen global magnitude limit. 
The
  dashed portion allows for compact groups containing isolated compact subgroups.
\label{flujo}}
\end{figure} 
The main steps of this algorithm are summarised in the flowchart of
Figure~\ref{flujo}.

Now, some CGs meeting Hickson's criteria might be embedded within larger
CGs that also meet Hickson's criteria (with larger isolation annuli).
For such groups, we thus have two choices for our CG selection algorithm:
select the smaller (sub-)group (solid portion only in the flowchart of
Figure~\ref{flujo}) 
or the larger group (with the dashed portion of the flowchart of Figure~\ref{flujo}).
The percentages of CGs containing smaller CGs are  13\%, 10\% and 6\% for B06, C06 and DLB models, 
respectively.
The HCG sample was selected according to the larger group (P.~Hickson,
private communication).
However of the 100 groups in the original HCG sample, only one has a
definite subgroup (HCG~17).
Therefore, it is not clear that P. Hickson always followed the larger group algorithm.
In what follows, we adopt the larger group algorithm (i.e. including dashed
portion of the flowchart of Figure~\ref{flujo}). However, our results turn out
to depend little on the choice among these two algorithms.

To accelerate this 
algorithm, we have used the subroutines of the 
HEALPix\footnote{http://healpix.jpl.nasa.gov/index.shtml} package to find neighbours, 
and the STRIPACK\footnote{http://people.sc.fsu.edu/\~{}burkardt/f\_src/stripack/stripack.html}
subroutines to compute the centres and radii of the minimum circles.
Given that our mock catalogues have edges (the limits of the cone), we
discarded CGs lying near the edges 
since those groups will be fictitiously isolated. 
Then, we kept with a safe sample of CGs that lies in the range
$\alpha>5^\circ$ \& $\alpha<85^\circ$,  
and $\delta>5^\circ$ \& $\delta<85^\circ$ (solid angle $\Delta
\Omega=1.2693\,\rm sr$).

Using this algorithm, we find
7580, 4756 and 15383 mock CGs in the B06, C06 and
 DLB samples, respectively.

Now, the galaxies in the mock galaxy catalogues are simply point particles.
However, when one observes two galaxies that lie so close in projection on the
plane of the sky  that
their isophotes overlap, they risk being blended into a single object.
This galaxy confusion can be important for CGs, which by definition often have
overlapping isophotes. For example, observed CG catalogues should have fewer
very dense groups than mock CG catalogues.
We therefore included one extra \emph{observability} criterion: two galaxies 
are confused and blended if their projected separation is smaller than the
sum of their half light radii, in which case we sum their luminosities and adopt
the redshift of the most luminous galaxy.
For galaxies in the mock catalogues, the half light radii were computed as a function of their
absolute magnitude in the R-band, according to \citeauthor{Shen03}
(\citeyear{Shen03}, eqs. [14] and [15] therein).

Hereafter (see Table~\ref{acronyms}), we refer to these observable mock
  projected compact groups as 
  \emph{mpCG}s and denote the original ones as \emph{pmpCG}s (for
  \emph{particle-mock-projected compact groups}).
The \emph{mpCG}s are built with the Hickson criteria given at the beginning
of Sect.~\ref{criteria} and thus
contain at least 4 galaxies (after the pair-blending procedure).
We understand that our pair-blending criterion is simplistic and may be
somewhat liberal in defining confused 
galaxy pairs. In reality, observed CGs should lie in between our \emph{pmpCG}s and
our \emph{mpCG}s, but probably much closer to the \emph{mpCG}s.
We therefore adopt the observable criterion (hence, the \emph{mpCG}s) in what follows, unless
explicitly stated otherwise.

\subsection{Mock compact groups after velocity filtering}
\label{redshift}

We then built a sample of \emph{velocity-filtered mock compact groups} on top
of the respective \emph{pmpCG} and \emph{mpCG} samples, which we call
the \emph{pmvCG} (\emph{particle-mock-velocity-filtered compact group}) and
\emph{mvCG} (\emph{mock velocity-filtered compact group}) samples (see
Table~\ref{acronyms}) with the following 
iterative procedure (see \citealp{HMdOHP92}):
\begin{itemize}
\item Compute the median velocity of the group, $v_{\rm median}$.
\item Discard those galaxies with $|v-v_{\rm median}|>1000 \rm\,km\,s^{-1}$.
\item If at least $n_{\rm min}$ galaxies remain, 
iterate until no galaxies are dropped or the group disappears ($n<n_{\rm min}$),
\item Save those CGs that have at least $n_{\rm min}$ galaxies and that satisfy the
  compactness criterion. 
\end{itemize}
We call $n$ the number of accordant-velocity galaxies in the \emph{mvCG} and
adopt $n_{\rm min}=4$ as our minimum number of accordant velocities. 

Table~\ref{samples} shows the number of groups in the observed and mock CG samples.
\begin{table}
\begin{center}
\tabcolsep 5pt
\caption{Compact group samples \label{samples}}
\tabcolsep 2pt
\begin{tabular}{rcccccccccc}
\hline
Sample & Type & gals & $v$
&
\multicolumn{3}{c}{$N$}
& & \multicolumn{3}{c}{$\overline n_{90}$} \\
& &  
 & filter
& & & & & \multicolumn{3}{c}{($10^{-5} h^3\rm Mpc^{-3}$)}  \\
\hline
\emph{pHCG} & obs & ext & no & 
\multicolumn{3}{c}{\ \ 72} & &
\multicolumn{3}{c}{\ \ \ 1.9} \\ 
\emph{vHCG} &  obs & ext & yes & 
\multicolumn{3}{c}{\ \ 52} & &
\multicolumn{3}{c}{\ \ \ 1.1} \\ 
& & & &
B06 & C06 & DLB & & B06 & C06 & DLB  \\
\cline{5-7}
\cline{9-11}
\emph{pmpCG} & mock & par & no & 
7580 & 4756 &  15383 & &  49\ \ & 33\ \ & 31\ \ \\ 
\emph{mpCG} & mock & ext & no & 
3574 & 3265 & \ \,4729 & &  18\ \ & 23\ \ &  11\ \ \\ 
\emph{pmvCG} & mock & par & yes & 
4553 & 2685 &  \ \,5646 & &  43\ \ & 29\ \ & 23\ \ \\ 
\emph{mvCG} & mock & ext & yes & 
2073 & 2095 &  \ \,2825 & &  16\ \ & 22\ \ & 10\ \ \\ 
\emph{mvHCG} & mock & ext & yes & 
\ \,272 & \, 223 &  \ \ \ 291 & & \ \ \ 5.1 & \ \ \ 5.4 & \ \ \ 2.6  \\ 
\hline
\end{tabular}
\end{center}
\parbox{\hsize}{Notes: 
Col. 1: sample;
col. 2: sample type (obs=observed);
col. 3: galaxy type (par=particle; ext=extended);
col. 4: velocity filter;
col(s). 5: number of groups (summed over 8 vertexes for mocks);
col(s). 6: space density within $90 \, h^{-1} \, \rm Mpc$ (eq.~[\ref{n90}],
 divided by 8 for the mock samples
to take into account
the 8 vertexes from which they were selected).
Columns with 3 values show the results for the 
\citeauthor{Bower+06} (B06), \citeauthor{Croton+06} (C06), and
\citeauthor{dLB07} (DLB) SAMs,
respectively. 
The limiting surface magnitude of the \emph{mvHCG} sample is not
sharp. 
}
\end{table}
%
The percentage of \emph{mpCG}s that survive the velocity-filtering is 
58\% (B06), 
64\% (C06), and 
60\% (DLB), 
so that our final samples 
of accordant velocity CGs contain from $\sim 2050$ to $\sim 2800$ \emph{mvCG}s 
depending on the adopted SAM. 

\begin{figure}
\centering
\resizebox{\hsize}{!}
{\includegraphics{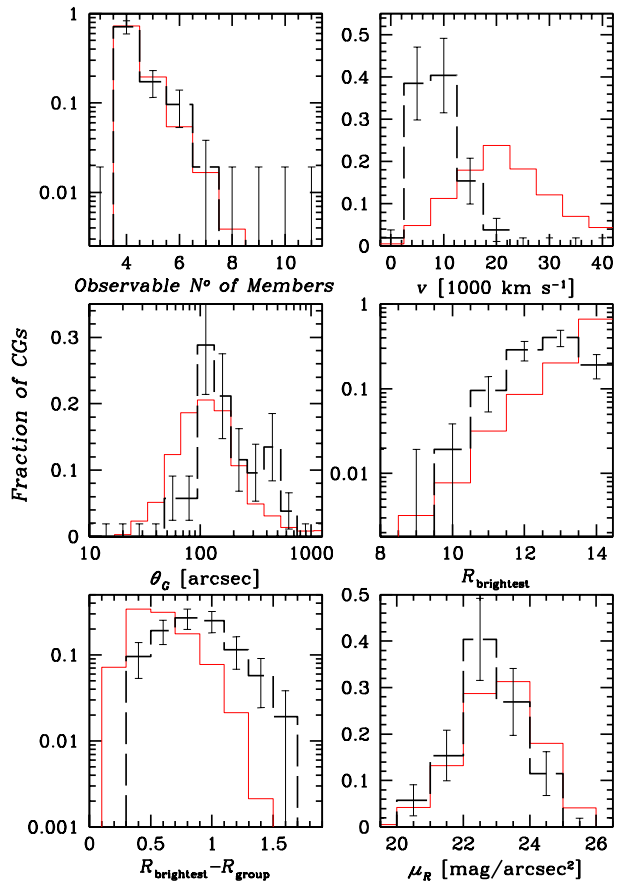}}
\caption{ 
Distributions of properties of the velocity selected CGs:
 mock
\emph{mvCG}s from the De Lucia \& Blaizot (2007) model (\emph{thin red
  histograms}) and observed \emph{vHCG}s
(\emph{thick black dashed histograms}).
The properties are the group
multiplicities (\emph{top left panel}), 
radial velocities (\emph{top right panel}),
angular group diameters (\emph{middle left panel}),
brightest galaxy magnitudes (\emph{middle right panel}),
magnitude differences between the brightest galaxy and the full
group (\emph{bottom left panel}),  and
group surface brightness (\emph{bottom right panel}). 
Error bars correspond to Poisson errors.
\label{distribs}}
\end{figure}

The main properties of the \emph{mvCG}s identified in the
DLB 
galaxy catalogue are shown in
Figure~\ref{distribs} together with the observed distribution of \emph{vHCG}s (the
distribution of the \emph{mvCG}s obtained with the other two SAMs are
similar, except for radial velocity distributions that are more skewed to
lower values and considerably more groups in the bin of lowest group surface brightness).
Hickson's visual selection of CGs produced a catalogue that is incomplete at 
small angular sizes (middle left plot),
faint brightest galaxy magnitudes (middle right plot), and
in groups with a dominant brightest galaxy (bottom left plot).
We will quantify the completeness of the HCG in Sect.~\ref{complete}.

\subsection{Testing the volume-limited sample}
\label{testing}

As we shall now see, limiting the depth of our galaxy sample to the
simulation box provides a complete list of \emph{mvCG} candidates. However, our
neglect of galaxies further than the box size may prevent distant galaxies
from spoiling the isolation of some of the \emph{mvCG}s.
Although the DLB model is available in an observing cone, this is not the
case for the other two SAMs,
and while we can construct a cone ourselves by
placing galaxies of previous time-steps at the position corresponding to
 their lookback times, we do not have access to the $z>0$ outputs of the
\cite{Croton+06} model to do this, and we wish to consider all three SAMs in
parallel. 

We therefore use the DLB model to build a mock sample of galaxies within a
cone, built of shells constructed from different snapshots corresponding
to the epoch of the lookback time at their distance. Here we use the 17 last
snapshots, bringing us to a maximum redshift of $z=0.68$, where the minimum
luminosity, $M_R = -24.91$, corresponds to $45\,L_*$.

\begin{figure}
\centering
\resizebox{\hsize}{!}{\includegraphics{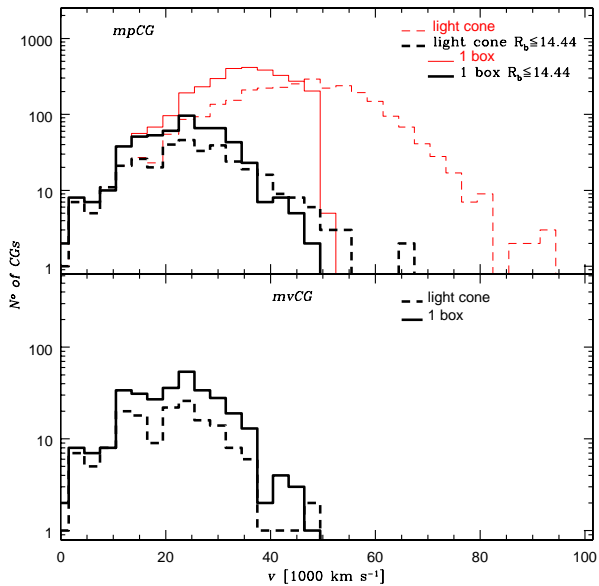}}
\caption{Radial velocity distributions of \emph{mpCG}s (top panel) and
  \emph{mvCG}s (bottom panel) built from a volume-limited catalogue (solid lines) 
and from a light cone (dashed lines), both having $R<17.44$. These catalogues were
  constructed from the DLB SAM. 
\label{test}}
\end{figure}

The top panel in Fig.~\ref{test} compares the radial velocity distribution 
of \emph{mpCG}s extracted from
the volume limited mock catalogue described in Sect.~\ref{mockgals} 
(\emph{solid lines}) seen from one of its vertexes with that obtained 
from groups identified in a magnitude limited mock catalogue or light cone 
of the same solid angle
(\emph{dashed lines}).
It can be clearly seen that the \emph{mpCG}s obtained from the light cone and from 
the volume limited catalogue before the $R_{\rm brightest}$ cut-off (thin lines) 
are quite different. 
First, the cone sample is able to detect a large number of \emph{mpCG}s beyond the
limits of the box. On the other hand, fewer \emph{mpCG}s are identified in the cone
sample at small distances (up to 80\% of the box size). This lower abundance
of \emph{mpCG}s in the cone sample is the consequence of distant galaxies spoiling
the isolation criterion of many mock compact groups.
The total number of \emph{mpCG}s in the cone sample is 4\%
lower than in the box sample.
These differences become more pronounced 
when the flux limit is applied (thick lines).
The number of \emph{mpCG}s in the cone sample is now 38\% lower 
than that of the box sample.
Interestingly, the cone sample of \emph{mvCG}s (with the brightest galaxy magnitude
limit applied) shows a lack of CGs at all distances in comparison with
the analogous box sample. The total number of \emph{mvCG}s in the cone sample is half
that of the analogous box sample.
This indicates that our \emph{mpCG} box-sample catalogue is only 62\% reliable
against contamination of the isolation annulus by distant interlopers, while
our \emph{mvCG} box-sample 
catalogue is only 50\% reliable. 

We can use this comparison of box and cone samples to correct the fraction of
\emph{mpCG}s that survive the velocity filter and make it as \emph{mvCG}s: 
this fraction becomes $0.6\times 0.5/0.62=48\%\pm1$ (where the error is from
binomial statistics and neglects the systematic error from the cone to box
correction). 


\subsection{Completeness}
\label{complete}
\subsubsection{Measure of completeness}

It is interesting to compare the space density of \emph{mpCG}s and \emph{mvCG}s
with those of the observed HCGs, selected in the same way.
We estimate for the mock and observed samples the mean total surface density of
CGs, as well as the mean space density of CGs
within the lowest median distance of all samples, which is a fairly robust measure of
density. The adopted distance is $9000 \, \rm km \, s^{-1}$, 
which is close to the median of the \emph{vHCG} sample,
so the space density
is computed as
\begin{equation} 
\overline n_{90} = {3\,N\left(v<9000 \, {\rm km \, s^{-1}}\right) \over
90^3\,\Delta \Omega} \, h^3\,{\rm Mpc}^{-3} \ ,
\label{n90}
\end{equation}
where $v$ is the median velocity of the group members, while $\Delta$ is the
solid angle of the sample (in sr).

\subsubsection{Projected compact groups}

For the \emph{mpCG}s, we obtain  
1.8, 2.3 and $1.1 \times 10^{-4} \,h^3\,\rm Mpc^{-3}$ (see Table~\ref{samples})
using the SAMs by 
B06, C06 and DLB, respectively. 
In comparison, the HCGs were selected
on the POSS~I plates, spanning $9.7\,\rm sr = 32\,000\,deg^2$ ($\rm Dec >
-33^\circ$). 
For the 72 \emph{pHCG}s,
the mean density is
$\overline n_{90} = 1.9 \times10^{-5}\,h^3\,\rm Mpc^{-3}$
(Table~\ref{samples}), i.e. typically 9 
times lower than the 
values obtained  from the 3 samples of \emph{mpCG}s.

Now, within a limiting distance of $v=9000 \, \rm km \, s^{-1}$, 
we found (Fig.~\ref{test}) 17 \emph{mpCG}s in our light cone in comparison with 
22 in one of our boxes, again because our box sample misses possible distant
interlopers that spoil the CG isolation.
This suggests that we would have found 23\% fewer
\emph{mpCG}s, 
had we not limited ourselves to the box.
We deduce that the observed \emph{pHCG} sample is $ (1/9) / (17/22) \sim 14\%$ 
complete at this limiting
distance (which again corresponds to the median distance of the HCG catalogue).

Hickson's inclusion of the Galactic Plane should lead to underestimates of
the completeness of roughly 1/3, which is the fraction of his search area
($\delta > -27^\circ$ covered by the POSS~I survey)
with low galactic latitudes $|b| < 20^\circ$.
Therefore, the bulk of the incompleteness of the HCGs lies in the incomplete
visual selection at high galactic latitudes.

\subsubsection{Velocity-filtered compact groups}

We now compare the space density of \emph{mvCG}s 
with that of the \emph{vHCG} sample.
For our mock samples of
\emph{mvCG}s with at least 4 accordant velocities, the space densities
$\overline n_{90}$ are (Table~\ref{samples})
1.6, 2.2 and $1.0 \times 10^{-4} \,h^3 \,\rm Mpc^{-3}$, for
B06, C06 and DLB, respectively.
For comparison, for the 52 \emph{vHCG}s (defined with at least 4
accordant velocities and with brightest galaxy magnitude brighter than 14.44), 
the space density is
$\overline n_{90} = 1.1 \times 10^{-5} \,h^3 \,\rm Mpc^{-3}$
(Table~\ref{samples}).
Therefore, the space density of \emph{mvCG}s selected in the box
  is typically 15 times that of the observed \emph{vHCG}.

However, within $v < 9000 \, \rm km \, s^{-1}$, we found
16 \emph{mvCG}s in our light cone versus 20 (20\% more)
in our box (for a single vertex as observation point). 
This suggests that we would have found 20\% fewer
\emph{mvCG}s, had we not limited ourselves to the box (thus allowing for
distant galaxies to spoil the isolation of these 20\% of the \emph{mvCG}s).
Therefore, we deduce that the completeness of the \emph{vHCG} sample is $1/15
/ (16/20) = 8\%$.
Note that we assumed that the contamination of distant galaxies of the
isolation criterion of \emph{mpCG}s and \emph{mvCG}s is independent of the
SAM, even if we only measured this effect with the DLB model.

\begin{figure}
\centering

{\includegraphics[scale=0.5]{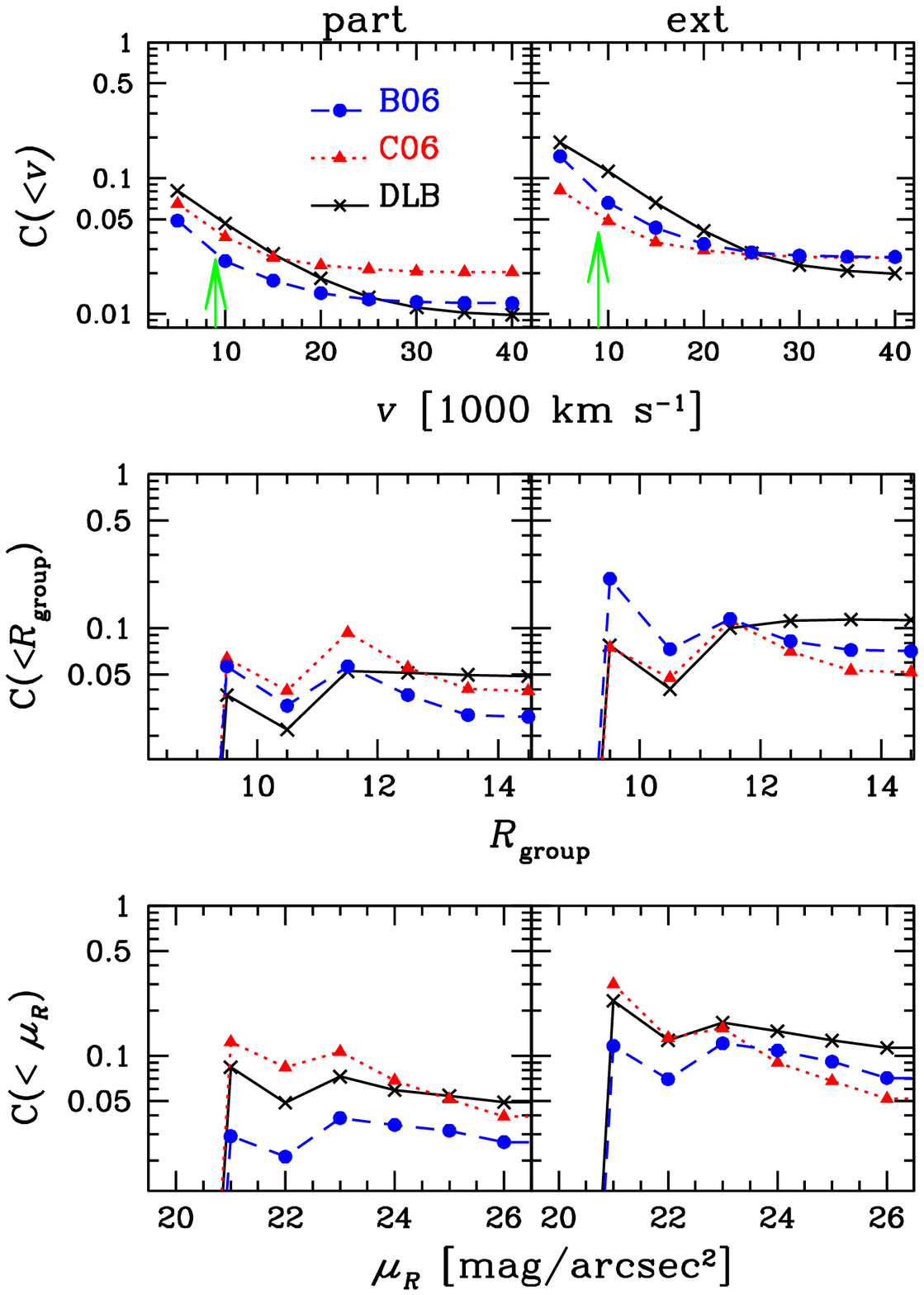}}
{\includegraphics[scale=0.5]{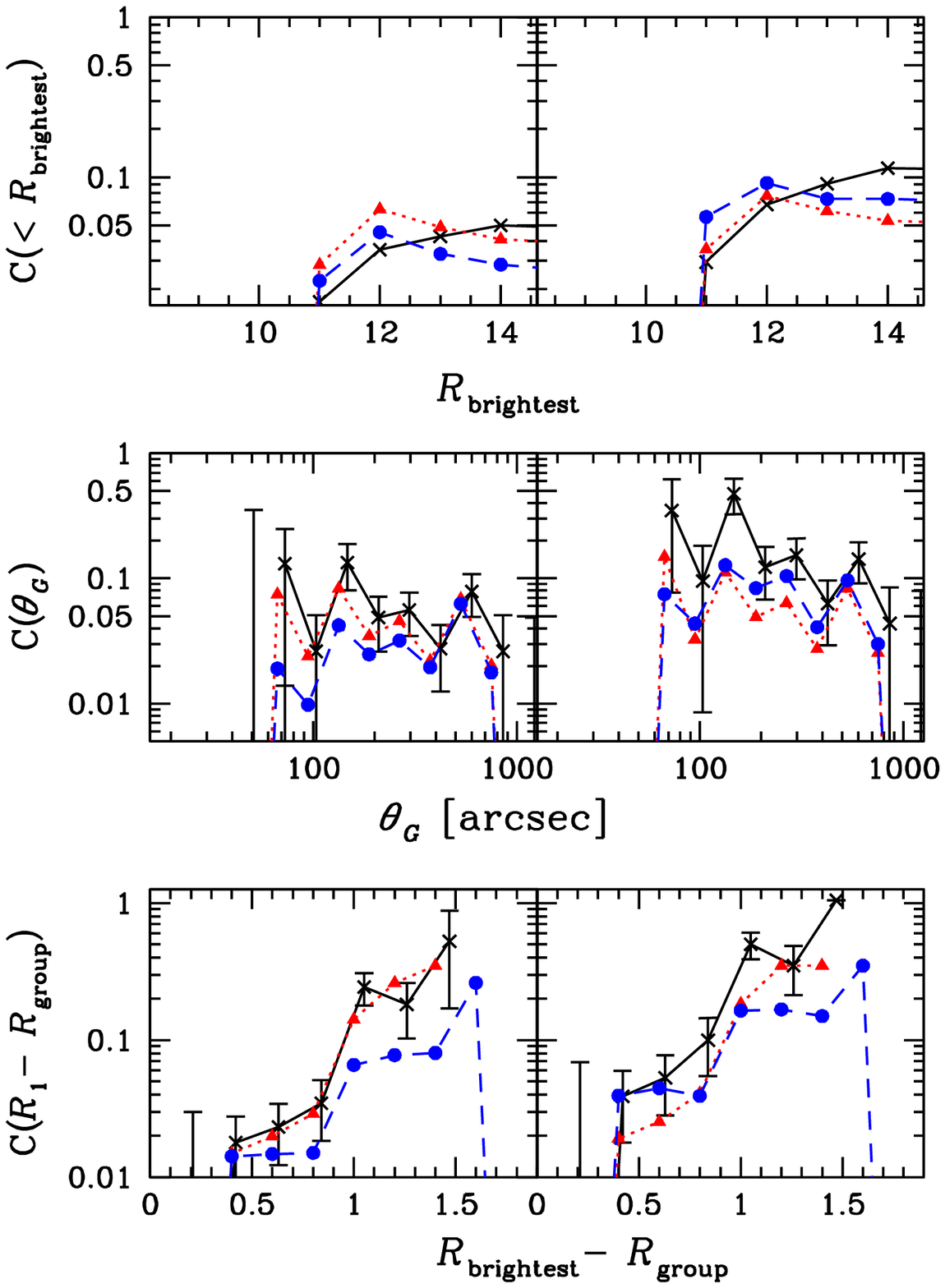}}

\caption{Completeness of the \emph{vHCG} catalogue relative to the \emph{pmvCG}
  (\emph{left}) and \emph{mvCG} (\emph{right}) mock catalogues as a function of
  distance (\emph{top panels}) and for groups within $v=9000 \, \rm km \,
  s^{-1}$ (\emph{other five rows of panels}).
From top to bottom the panels represent
global cumulative completeness vs. distance, vs. group total magnitude, 
vs. group surface brightness and vs. brightest group galaxy, and
differential completeness vs. angular size and
vs. dominance of brightest galaxy.
The reference SAMs are 
Bower et al. (\emph{blue circles}),
Croton et al. (\emph{red triangles}),
and
De Lucia \& Blaizot (\emph{black crosses}). Error bars for differential completeness 
are computed as binomial errors and are shown only for DLB SAM.
\label{completeness}}
\end{figure}

The top panels of Fig.~\ref{completeness} show the completeness of the
velocity-filtered Hickson sample as a function of radial velocity for the 3
SAMs. The completeness is defined as $C(v) = \overline n_{v/H_0}^{vHCG}/\overline
n_{v/H_0}^{mvCG}$, where $H_0 = 100 \,\rm km \,s^{-1} \, Mpc^{-1}$.
The green arrow shows the limit of our nearby subsample (see
Table~\ref{samples}). 
The next 5 rows of panels show
the completeness within the nearby subsamples ($v < 9000 \, \rm km \, s^{-1}$)
$C=\overline n_{90}^{vHCG}/\overline
n_{90}^{mvCG}$,
as a function of the other observable properties for the
nearby subsample.

For all 3 SAMs, the \emph{vHCG} completeness decreases sharply with distance
(top panel),
for groups with a dominant brightest galaxy (bottom panel), while the C06 model also
predicts 
a decrease of \emph{vHCG} completeness at fainter magnitude and 
lower surface brightness, whereas
these trends are weaker in B06 and absent in DLB.
Comparing the left and right sets of panels, one sees similar qualitative
trends of completeness versus interesting parameter, but with overall completeness
relative to the extended-galaxy \emph{mvCG}s that is
roughly double the value of the completeness relative to
particle-based \emph{pmvCG}s.
Moreover, one can see a slightly faster trend of decreasing completeness with
decreasing surface brightness with the C06 model.

\subsubsection{Comparison with previous studies}

Previous studies have concluded that the \emph{pHCG} sample is incomplete at low group surface
brightness ($24 < \mu_R < 26\,\rm mag\,arcsec^{-2}$):
\cite{Hickson82} and \cite{WM89} (from the lack of low surface brightness
groups), and \cite*{PIM94} (from a comparison with their own
automatically-selected projected SCG sample of compact groups, built 
with very similar criteria as \citealp{Hickson82}).
We also found a large incompleteness at low group surface brightness when comparing
the \emph{pHCG}s with our \emph{mpCG} samples. 
However, the sharp drop in the number of compact groups with low surface
brightness ($\mu_R > 24$) observed in the \emph{pHCG}s, 
is also clearly visible in the \emph{mvCG}s with the DLB
model (but less so with the other two SAMs).

The incompleteness in brightest galaxy counts 
is analogous to the incompleteness in
group number counts that \citeauthor{Hickson82} had noticed at $R=13.0$ and
that \citeauthor{PIM94} had already noticed at the
much brighter limit of $b_J = 13.1$ (from the break in the slope of the
number counts away from the Euclidean value of 0.6). 
\cite{Mamon00_IAU174} noted that 
Fig. 7 of \citeauthor{PIM94} 
indicates that the \emph{pHCG} catalogue is incomplete
by a factor 3 at bright magnitudes, relative to the SCG catalogue,
while this incompleteness gets worse at increasingly fainter magnitudes.
A closer look at their Fig. 7 reveals that the number of groups brighter than
$b_J = 13$
is roughly 30 for the HCG and 2.5 for the Euclidean 
extrapolation of the SCG group counts to
this relatively bright magnitude. Given that the solid angle of the HCG
($\rm 32\,000\,deg^2$) is 25 times that of the SCG ($1300\,\rm deg^2$), the
completeness of the HCG relative to the SCG is $30/2.5/25 = 0.48$, with
total surface densities of $30/32\,000 = 0.9\times 10^{-3} \,\rm deg^{-2}$ and
$2.5/1300 = 1.9\times10^{-3}\,\rm deg^{-2}$ for the \emph{pHCG} and SCG, respectively.

This strong incompleteness at faint magnitudes 
is also evident for \emph{vHCG}s, as seen 
in the middle right panel of our Fig.~\ref{distribs}, which
suggests (by matching the magnitude counts at intermediate magnitudes) 
that the HCG becomes incomplete for brightest galaxy magnitudes fainter than
$R=12.5$, to the point where at $R=14.44$, the differential completeness falls
to roughly 5\%.
The surface densities of the \emph{mpCG}s limited to 
magnitude brighter than $R<11.6$ (roughly corresponding to $b_J = 13$) are
${7.5, 6.6, 11.8}\,\times10^{-3}\rm deg^{-2}$, for B06, C06 and DLB,
respectively, typically 10 times that the surface density of the \emph{pHCG}s.
However, once
we limit groups to the nearby subsample, 
the strong incompleteness at faint magnitudes appears barely visible (B06 and C06)
or reversed (DLB), as seen in Figure~\ref{completeness}.

The incompleteness of the \emph{pHCG} in dominant brightest galaxy groups had already
been 
noticed by \cite{PIM94}, who also found that \cite{Hickson82} 
was biased in favour of
groups where the two brightest galaxies have comparable magnitudes.

\subsection{Mock Hickson compact groups}
\label{mock}
As noted above,
the HCGs produced by
Hickson's visual inspection cannot be reproduced by an 
automatic searching algorithm given the many biases in the selection of HCGs.
Therefore, the nature and properties of the
mock CGs that strictly meet the HCG criteria mentioned in
Sect.~\ref{criteria} may 
be different from the properties of the HCGs themselves.

Given the strong and progressive incompleteness of the HCGs 
in brightest galaxy counts, small angular sizes, and systems with strongly
dominant brightest galaxies, it is essential to fold in these extra factors
of incompleteness 
when building a sample that will be a good mock for the observed HCGs.
We therefore wish to construct a mock velocity-selected Hickson Compact Group
(\emph{mvHCG})  sample, starting with the \emph{mvCG} sample, and selecting
galaxies with probabilities proportional to the completeness in 1) group surface
brightness, 2) brightest galaxy magnitude,
and 3) difference between the brightest galaxy and total group magnitude
(i.e. the relative importance of the brightest galaxy).
We do not consider the distribution of angular sizes, since this latter
quantity is directly dependent on the three other parameters.

Because the resulting number of \emph{mvHCG}s turns out to be very small,
rather than select \emph{mvCG}s according to the probability that a given
\emph{mvCG} would be observed by Hickson, we proceeded as follows.
We selected the first \emph{mvCG}s
that fill the observed distribution of \emph{vHCG}s for the three parameters
and stopped once one of the 10 bins in any of the three distributions for the
\emph{mvCG}s
reaches the
value observed in the corresponding bin for the \emph{vHCG}s.
Hence, the derived distributions of the three parameters do not match
perfectly the observed ones, but are lower limits.
We repeated this exercise, using different orders for our loop over the
\emph{mvCG}s until we matched as best as possible the observed \emph{vHCG}
distributions.
\begin{figure}
\centering
\resizebox{\hsize}{!}{\includegraphics{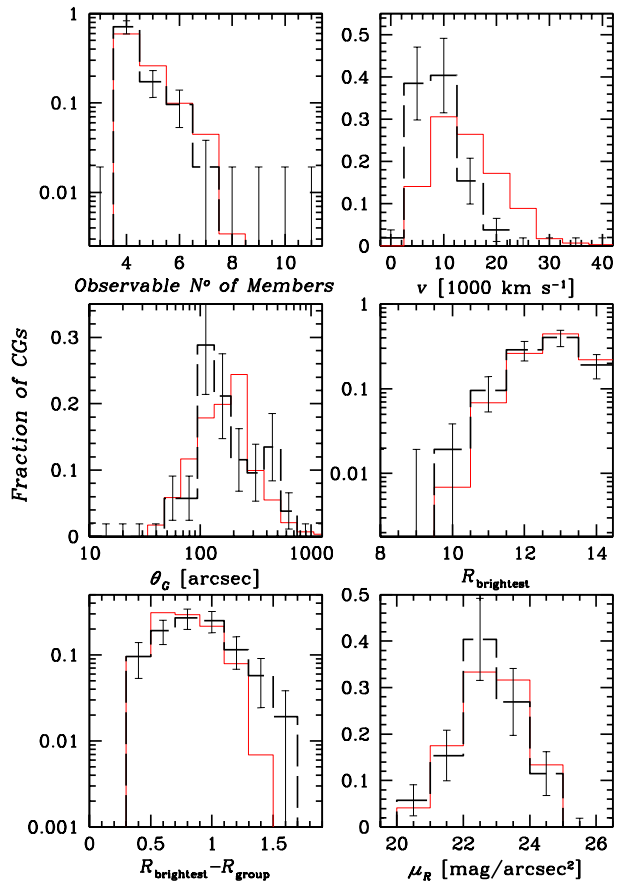}}
\caption{ Same as Figure~\ref{distribs} for the \emph{mvHCG}s 
from the \citeauthor{dLB07} SAM (\emph{thin red histograms}) 
and (again) the observed \emph{vHCG}s (\emph{thick black dashed
  histograms}). 
\label{random}
}
\end{figure}
This procedure was again applied on the eight different samples corresponding to the 
eight observers situated at the eight vertexes of the simulation cube 
obtaining final samples of typically 250 \emph{mvHCG}s. 

The distribution of properties of \emph{mvHCG}s, shown in Figure
\ref{random} for the DLB model, matches much better the observed distributions of \emph{vHCG}s.
Similar results are found using the other two SAMs. 
The three SAMs fail to find groups of roughly concordant magnitudes.

The three \emph{mvHCG} samples  
will be used to compare with the properties of observed \emph{vHCG}s and with 
the correlations obtained for previous authors based on the observed
accordant-velocity HCGs.

\section{Different classes of Compact Groups}
\label{classes}
Even though we have used redshift information to identify our \emph{mvCG}s, 
the selected groups are not necessarily physically dense in 3D real space.

Our \emph{mvCG}s can thus be split
 into three classes:
\begin{itemize}
\item {Physically dense groups (Real)},
\item {Chance alignments within loose groups (CALG)},
\item {Chance alignments within filaments (CAF)}.
\end{itemize}
Keeping with the original intent of \cite{Hickson82}, who had selected in
projection compact groups of at least four galaxies, we will classify an \emph{mvCG}
as Real if at least 4 of its galaxies form a physically dense group.
Also, we will sometimes join the CALG and CAF classes into the set of Chance
Alignments (CAs).

There are several ways to use the three-dimensional information to define
these classes, and as we shall see, none of them are perfect.

\subsection{Binding energies}

A simple way to separate the Real CGs from the CAs is to
use the binding energy of the system. In appendix~\ref{appbind}, we show that
binding energies are highly inaccurate for groups of masses
$M<10^{14}\,h^{-1}\,M_\odot$, and cannot be used to distinguish which CGs
are physically
dense and which are caused by chance alignments.

\subsection{Line-of-sight shape and 3D length}

Alternatively, we can classify the mock CGs using their size and/or
 elongation.
We consider the 4 closest galaxies in each mock CG, again in line
with the original intent of \cite{Hickson82}. By closest 4 galaxies, we mean
either the entire \emph{mvCG} if it has only 4 galaxies, or else the subgroup
of 4 with
the smallest 3D length.
We use the following
notations for these smallest quartets:
\begin{itemize}
\item $s$: maximum 3D separation, hereafter \emph{3D length};
\item $S_\perp$: maximum projected separation, hereafter \emph{projected size};
\item $S_\parallel$: maximum line-of-sight separation, hereafter
  \emph{line-of-sight length};
\item $S_\parallel/S_\perp$: hereafter, \emph{line-of-sight elongation};
\item \emph{round} \emph{mvCG}: $S_\parallel/S_\perp < 2$;
\item \emph{elongated} \emph{mvCG}: $S_\parallel/S_\perp \geq 2$.
\end{itemize}

\begin{figure}
\centering
\includegraphics[width=\hsize]{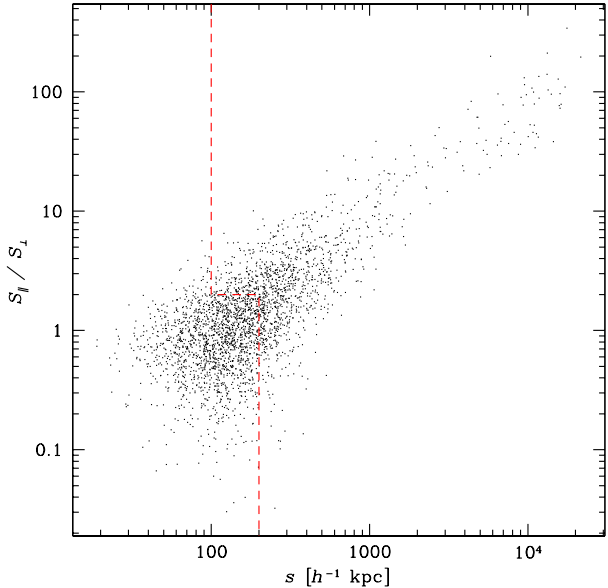}
\caption{Line-of-sight elongation vs. maximum 3D separation of mock
  velocity-filtered compact
  groups extracted from the De Lucia \& Blaizot (2007) model.
The \emph{red dashes} delimit the Real \emph{mvCG}s with the hybrid classification
discussed below.
}
\label{Sratvss}
\end{figure}

Figure~\ref{Sratvss} shows how the line-of-sight elongation is related to the
maximum 3D separation.
The points in the lower left part of Figure~\ref{Sratvss} show uncorrelated
line-of-sight elongation and 3D length, as expected for Real groups, while
the upper right part of the figure shows instead a strong correlation of
line-of-sight elongation with 3D length, indicative of CA groups.
Which cuts in line-of-sight elongation and 3D  length separate best the Real
CGs from the CAs?
The choice of the critical 3D length, $s_{\rm cut}$, is not straightforward,
as we shall now
discuss.

\subsubsection{Matching line-of-sight elongation of real-space selected
  groups}

We first tried varying $s_{\rm cut}$ by imposing that the
median line-of-sight elongation be equal to that of real-space-selected
groups. We measured a median line-of-sight elongation of 0.725 for the PG3Ds.
We also
checked this median value of the line-of-sight elongation with
Monte-Carlo simulations of quartets distributed at random in a virial sphere with an NFW
density profile with concentration $r_v/r_s=10$, where we then elongated the sphere in
two orthogonal directions by two factors to make it a triaxial ellipsoid, 
and observed it from a random direction, and repeated this
exercise 5000 times. We then find median line-of-sight elongations of 0.782
(in spheres) and 0.719 (in triaxial ellipsoids, with $b/a=0.79$ and $c/a =
0.65$, as found by \cite{JS02}, on average, in $\Lambda$CDM halos at
overdensity 100). 
This median line-of-sight elongation for triaxial halos is very close to what
we measured for the PG3Ds.

We also considered the cores of virialized groups, where the overdensity is $10^5$,
close to how overdense HCGs appear to be. Here we, limited the particles to a
radius of 0.025 virial radii, i.e. 0.25 scale radii (with our concentration
of 10), where the mean density
is roughly 1000 times greater than at the virial radius.
Noting that $\Lambda$CDM halos are less spherical at overdensities as high as
$10^5$ (\citeauthor{JS02} find $b/a=0.61$ and $c/a=0.46$, on average), we
consider these less spherical halos and then find a median line-of-sight
elongation of 0.722 (even closer to the median elongation of the PG3Ds).
In general, the median line-of-sight elongation is much more sensitive to the
triaxiality of the object than to the slope of its density profile.

\begin{figure}
\centering
\includegraphics[width=\hsize]{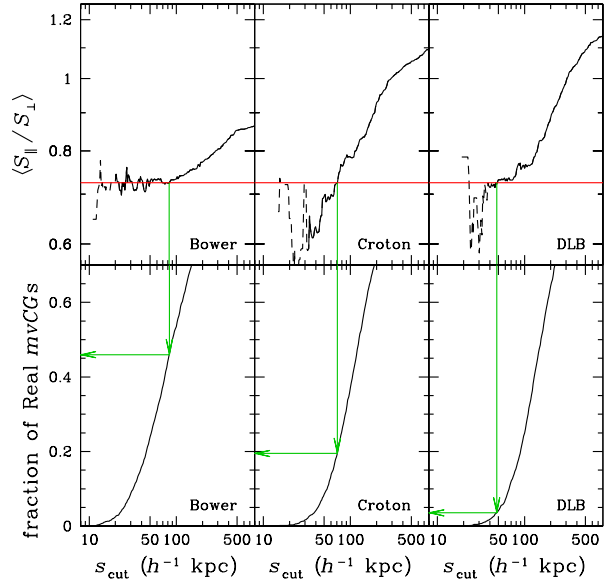}
\caption{\emph{Upper panels}: Median 
  line-of-sight elongation vs. critical 3D length,  $s_{\rm cut}$,
  for the 4 galaxies in the richest subclump of mock velocity-accordant
  compact groups. \emph{Dashed} (\emph{solid}) \emph{lines} refer to values of
  $s_{\rm cut}$ where the sample of \emph{mvCG}s has less (more) than
50 \emph{mvCG}s.
The \emph{red solid horizontal lines} correspond to the median line-of-sight
elongation measured by selecting 4 
galaxies at random from each \emph{PG3D} group. 
\emph{Lower panels}: 
Normalised cumulative distribution of 3D lengths, considered as critical
separations between the Real and CA classes.
The \emph{green arrows} indicate
  the value of $s_{\rm cut}$ that matches the median line-of-sight
  elongations of the PG3D groups (\emph{vertical arrows}) yielding the
  corresponding 
fraction of Real CGs 
for the adopted $s_{\rm cut}$ (\emph{horizontal arrows}).  
\label{rcut}}
\end{figure} 

As seen in the top panels of Figure~\ref{rcut}, 
the values of $s_{\rm cut}$
required for the shortest \emph{mvCG} to reproduce the median line-of-sight elongation 
of the real-space selected PG3Ds are fairly small and
vary from SAM to SAM, from $\sim 50 \, h^{-1} \, \rm kpc$ for DLB
to $\sim 80 \, h^{-1} \, \rm kpc $ for B06, with C06 in between.
The bottom panels of Figure~\ref{rcut} indicate that the fraction of \emph{mvCG}s
whose 3D length of their smallest subclump of four galaxies is less than the
corresponding $s_{\rm cut}$ varies strongly with the SAM: 40\% with B06, 20\%
with C06, but only 4\% with DLB.
This should not lead us to conclude that most \emph{mvCG}s are CAs, because the
minimum 3D length ($s_{\rm cut}$) for CAs with the DLB model is only half the
median projected size of the DLB \emph{mvCG}s, which does not seem reasonable.
In other words, it is not reasonable to force redshift-space selected groups 
to be as round as real-space selected ones: redshift-space selection will always produce
  somewhat more elongated groups than real-space selected ones.

\subsubsection{Line-of-sight elongation versus line-of-sight size}

Alternatively, one could argue that CAs should be long in the absolute,
i.e. high $s$, and/or relative to their projected sizes, i.e. high
$S_\parallel/S_\perp$, e.g.  $S_\parallel/S_\perp>2$.

If the projected sizes were independent of the line-of-sight
lengths, as would be expected if all CGs were CAs, 
we could then impose a value of $s_{\rm cut}$ that would be close to
$\sqrt{2}$ times the upper envelope of $S_\perp$ (since round CGs would have
$s\simeq \sqrt{S_\perp^2+S_\parallel^2} \simeq \sqrt{2}\,S_\perp$).

Figure~\ref{SperpvsSpar} shows that the \emph{mvCG}s behave differently:
\begin{figure}
\centering
\includegraphics[width=\hsize]{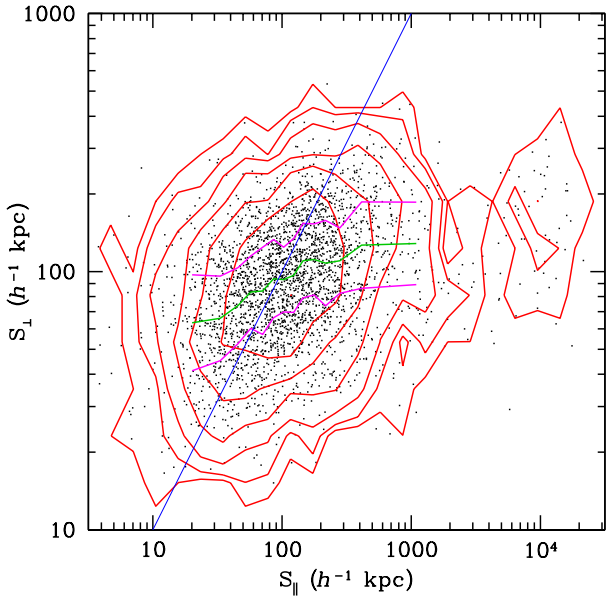}
\caption{Projected size versus line-of-sight length for the smallest quartets
  within \emph{mvCG}s, using the De Lucia \& Blaizot (2007) SAM.
Also shown are log-spaced contours (\emph{red}), the medians in bins of 200
points (\emph{green jagged line}) the interquartiles (\emph{magenta
  jagged lines}), and the line $y=x$ (\emph{thin blue line}).
}
\label{SperpvsSpar}
\end{figure}
while at high line-of-sight length, where CAs are expected to be dominant, the projected size
is indeed independent of the line-of-sight length, 
at low line-of-sight length, where CAs are not dominant, 
the projected size increases with increasing
line-of-sight length. So the
upper envelope of the 
projected sizes is not a clear-cut value.
The contours suggest a close to linear increase of $S_\perp$ with
$S_\parallel$ in the low $S_\parallel$ regime, as expected for 
systems of same line-of-sight elongations and different sizes.

The transition between these two regimes is difficult to ascertain.
One way is to look for the value of $S_\parallel$ for which the median
$S_\perp$ changes from a high to low slope.
This yields a critical $S_\parallel$ of $\approx 140 \, h^{-1} \, \rm kpc$ for
the DLB model (see Fig.~\ref{SperpvsSpar}), $120 \, h^{-1} \, \rm kpc$
for the B06 model and $165 \, h^{-1} \, \rm kpc$ for the C06 model.
These 3 critical values of $S_\parallel$ correspond to $S_\perp \simeq 100,
105$ and $112 \,
h^{-1} \, \rm kpc$, for the B06, C06 and DLB models, respectively. One
therefore infers critical group length of $s_{\rm cut} = 156, 196$ and $179 \,
h^{-1} \, \rm kpc$, for the B06, C06 and DLB models, respectively.

The fraction of \emph{mvCG}s with lengths smaller than these three values of
$s_{\rm cut}$ can then be
read from the bottom panel of Figure~\ref{sdis}: one finds 72\%, 72\% and
59\% of the groups have $s < s_{\rm cut}$ for the B06, C06 and DLB models,
respectively. One would therefore deduce that between half and three-quarters
of the \emph{mvCG}s are Real (depending on the SAM).
However, given the crudeness of the method, one should take these percentages
with caution.

\subsubsection{Reasonable cuts in length and line-of-sight elongation}

\begin{figure}
\centering
\includegraphics[width=\hsize]{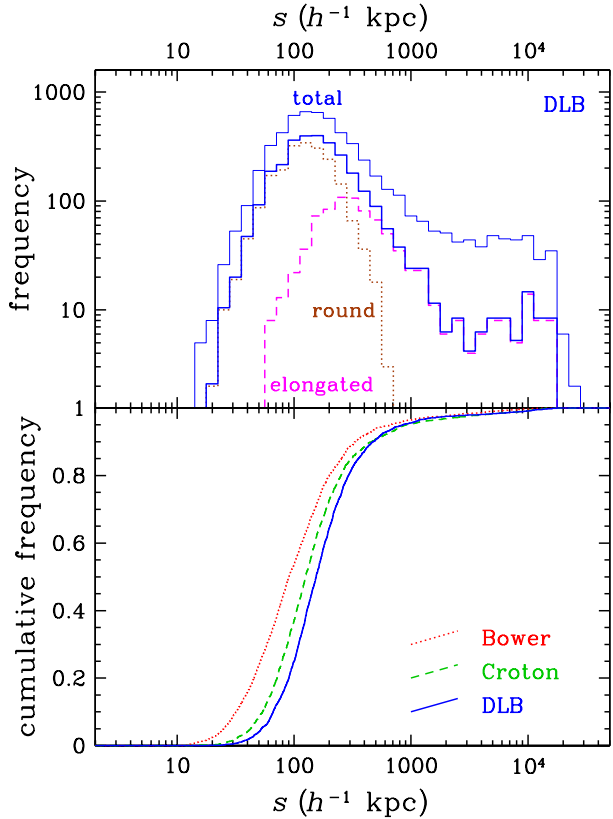}
\caption{\emph{Top:} Distribution of 3D lengths for the DLB model for
  all \emph{mvCG}s (\emph{solid histograms} pushed up by 20\% for clarity), 
the round \emph{mvCG}s
  ($S_\parallel/S_\perp < 2$, \emph{dashed histograms}), and
the elongated \emph{mvCG}s ($S_\parallel/S_\perp \geq 2$, \emph{dotted
  histograms}).
The \emph{thin solid blue curve} shows the total distribution for the \emph{pmvCG}s.
\emph{Bottom:} Cumulative distribution of 3D lengths of all \emph{mvCG}s.
\label{sdis}
}
\end{figure}

We now explore whether reasonable limits on $s_{\rm cut}$ and
$S_\parallel/S_\perp$ can reduce substantially the fraction of Real \emph{mvCG}s.
Figure~\ref{sdis} displays the distributions of the 3D length, $s$, for the
DLB model.

The distribution of 3D lengths clearly shows a dominant log-normal component and a second
component of more extended lengths. In fact, when restricting to round groups 
($S_\parallel/S_\perp < 2$), the distributions of 3D lengths appear very
close to lognormal.
Moreover, the round \emph{mvCG}s tend to be smaller with the Bower model
($\left \langle s\right\rangle = 78\, h^{-1} \, \rm kpc$)
and larger with the DLB model
($\left \langle s\right\rangle = 125\, h^{-1} \, \rm kpc$), 
with the predictions from the Croton model in
between. 
Finally, the distribution of elongated \emph{mvCG}s  ($S_\parallel/S_\perp \geq 2$) 
is wider than that of the round \emph{mvCG}s, and centred 
around $s \simeq 250 \, h^{-1} \, \rm kpc$ for the B06 and C06 models and $300
\, h^{-1} \, \rm kpc$ for the DLB model.
It displays an extended tail of very large ($> 2 \, h^{-1} \, \rm Mpc$) 3D lengths.

The discussion above suggests a conservative maximum for the Real \emph{mvCG}s of 
$s_{\rm cut} < 200 \, h^{-1} \, \rm
kpc$. 

\subsection{Fraction of CAs in different samples of mock CGs}

Table~\ref{nattab} summarises the fractions of CGs satisfying various
criteria that could classify them as Real.
With our choice of $s_{\rm cut} = 200 \, h^{-1} \, \rm kpc$,
we obtain fractions of Real \emph{mvCG}s of 
0.80 (Bower), 0.73 (Croton) and 0.64 (DLB).
To be more favourable to the CAs, we can include additional elongated
\emph{mvCG}s.
However, it makes no sense to call a CA an elongated \emph{mvCG} with a very small
3D length, for example with $s = 50 \, h^{-1} \, \rm kpc$, because such an
\emph{mvCG} is also a physically dense group, hence a Real.
So, we consider a simple hybrid classification (\emph{dashed lines} in
Fig.~\ref{Sratvss}), where the CAs are the \emph{mvCG}s 
with 
$s \geq 200 \, h^{-1} \, \rm kpc$
OR
$(s > 100 \, h^{-1} \, \rm kpc \hbox{ AND } S_\parallel/S_\perp\geq 2)$.
We then find (Table~\ref{nattab}) 
that the fraction of Real \emph{mvCG}s is 0.76 (Bower), 0.67 (Croton),
and 0.59 (DLB).
We hereafter adopt this hybrid criterion to estimate the fraction of
  Real and CA compact groups.

\begin{table}
\begin{center}
\caption{Fraction of Real mock CGs using different criteria
\label{nattab}
}
\begin{tabular}{lccc}
\hline
Criterion & B06 & C06 & DLB \\
\hline
\emph{pmpCG}\\
\hline
$s < 200 \, h^{-1} \, \rm kpc$ & 0.46 & 0.39 & 0.22 \\
$S_\parallel/S_\perp < 2$ & 0.43 & 0.38 & 0.23 \\
$s < 100 \, h^{-1} \, \rm kpc \hbox { OR}$ \\
$(s < 200 \, h^{-1} \, \rm kpc \hbox { AND } S_\parallel/S_\perp < 2)$ & 0.43
& 0.36 & 0.20 \\
\hline
\emph{mpCG}\\
\hline
$s < 200 \, h^{-1} \, \rm kpc$ & 0.46 & 0.47 & 0.38 \\
$S_\parallel/S_\perp < 2$ & 0.46 & 0.46 & 0.43 \\
$s < 100 \, h^{-1} \, \rm kpc \hbox { OR}$ \\
$(s < 200 \, h^{-1} \, \rm kpc \hbox { AND } S_\parallel/S_\perp < 2)$ & 0.44
& 0.43 & 0.35 \\
\hline
\emph{pmvCG}\\
\hline
$s < 200 \, h^{-1} \, \rm kpc$ & 0.77 & 0.70 & 0.59 \\
$S_\parallel/S_\perp < 2$ & 0.71 & 0.68 & 0.61 \\
$s < 100 \, h^{-1} \, \rm kpc \hbox { OR}$ \\
$(s < 200 \, h^{-1} \, \rm kpc \hbox { AND } S_\parallel/S_\perp < 2)$ & 0.72
& 0.64 & 0.53 \\
\hline
\emph{mvCG}\\
\hline
$s < 200 \, h^{-1} \, \rm kpc$ & 0.80 & 0.73 & 0.64 \\
$S_\parallel/S_\perp < 2$ & 0.80 & 0.72 & 0.71 \\
$s < 100 \, h^{-1} \, \rm kpc \hbox { OR}$ \\
$(s < 200 \, h^{-1} \, \rm kpc \hbox { AND } S_\parallel/S_\perp < 2)$ & 0.76
& 0.67 & 0.59 \\
\hline
\emph{mvHCG}\\
\hline
$s < 200 \, h^{-1} \, \rm kpc$ & 0.91 & 0.83 & 0.76 \\
$S_\parallel/S_\perp < 2$ & 0.84 & 0.73 & 0.76 \\
$s < 100 \, h^{-1} \, \rm kpc \hbox { OR}$ \\
$(s < 200 \, h^{-1} \, \rm kpc \hbox { AND } S_\parallel/S_\perp < 2)$ & 0.86
& 0.75 & 0.70 \\
\hline
\end{tabular}
\end{center}
\end{table}
It therefore, appears that \emph{more than half of the \emph{mvCG}s are physically
  dense}, although there are important variations between the three galaxy
formation models, with \citeauthor{Bower+06} predicting the most physically
dense \emph{mvCG}s, \citeauthor{dLB07} predicting the least, and
\citeauthor{Croton+06} in between. 

If we only consider CGs identified in projection (\emph{mpCG}s), 
the Real CGs represent between 35\% and 47\%, depending on the criteria and
on the SAM (Table~\ref{nattab}).

We note that the fraction of CAs diminishes in all three SAMs when going from
the \emph{pmvCG} to \emph{mvCG} and to \emph{mvHCG} samples, i.e. when
first taking into account the extended nature of galaxies causing confusion,
and then in incorporating the biases we measured in the visual selection of
\cite{Hickson82}. Table~\ref{nattab} thus indicates that the fraction of CAs
is 28--47\% for the \emph{pmvCG}s, but is reduced to 24--41\% for the
\emph{mvCG}s and only 14--30\% for the \emph{mvHCG}s.

\subsection{Chance alignments within Loose Groups and beyond}

We now consider as CAs all \emph{mvCG}s with 3D lengths $s \geq 200 \, h^{-1} \, \rm
kpc$, regardless of the line-of-sight elongation. 
The distinction between CALG and CAF is very simple, as we simply check
whether all CALG members lie within a single PG3D (making it a CALG) or not
(making it a CAF).
\begin{table}
\begin{center}
\caption{Fraction of chance alignments that extend beyond their parent loose group (CAFs)}
\label{CAs}
\begin{tabular}{lccc}
\hline
Sample & B06 & C06 & DLB \\
\hline
\emph{pmvCG} & 0.32 & 0.30 & 0.30 \\
\emph{mvCG} & 0.24 & 0.24 & 0.17 \\
\emph{mvHCG} & 0.16 & 0.26 & 0.26 \\
\hline
\end{tabular} 
\end{center}
\end{table}

Table~\ref{CAs} shows that large majority of the CAs (typically
  three-quarters) 
are CALGs, regardless of the SAM and the sample.
Thus, alignments within filaments or with galaxies in the field are much less
likely than chance alignments within larger groups or than having physically
dense groups.

\subsection{Isolated dense groups?}

The standard picture is that dense groups of galaxies are the cores of
virialized looser groups. However, 
some HCGs appear extremely isolated \citep{RW89,PSHM95}. 
The simulations that we have analysed allow us to check whether dense groups
can be isolated out to the virial radius.
We have simply cross-identified the \emph{mvCG}s with the real-space-selected
PG3Ds. 
We then call a CG  isolated 
if it constitutes the entire PG3D (in the magnitude range determined
by the brightest galaxy of the CG). 
We then find that only $\sim 11\%$ of the \emph{mvCG}s constitute the entire
PG3D in all the three SAMs.  
In the \emph{mvHCG} samples, the fraction of isolated CGs are somewhat
smaller 
(7\% for B06, 6\% for C06, and 8\% for DLB).

\section{Discussion and conclusions}
 \label{discus}
 \subsection{Summary of results}

 The aim of this paper is twofold: 1) predicting the nature of automatically 
 identified CGs, and 2) predicting the nature of the well-studied
 but highly incomplete and biased HCGs.
 We identify CGs in three mock galaxy catalogues, constructed 
 from the Millennium Simulation at $z=0$ combined with three semi-analytical
 models \citep{Bower+06,Croton+06,dLB07}.
 Several thousand mock 
 CGs are identified using a two dimensional automated algorithm similar 
 to that applied by \cite{Hickson82} plus a restriction in the brightest galaxy magnitude.
We also allowed for CGs that contain isolated and compact
subgroups and furthermore considered the important effects of extended galaxies
causing confusion for close projected pairs.

 Our main results are:
 \begin{enumerate}
 \item
Among the observable \emph{mpCG}s, 
$\sim 60\%$ have at least 4 
galaxies within $1000\,\rm km \, s^{-1}$ of the median velocity of the group, 
regardless of the SAM.
Since our study is carried out on a sample with a redshift cut-off ($z \sim 0.17$), 
we tested whether our results might be bias by this fact. We find that 
identifying in our (magnitude+)volume limited catalogue produces 1.6 times more \emph{mvCG}s than 
identifying on an only-magnitude limited catalogue, because 
our box sample misses distant galaxies that
spoil the isolation of mock Compact groups. Correcting for this effect, we
deduce that the fraction of \emph{mpCG}s that survive
the velocity filter is 50\%. In comparison, there are as many as 52
\emph{vHCG}s among 72 \emph{pHCG}s, meaning that 71\% of observed HCGs
survive the velocity filter. Given binomial statistics the probability that
we find as many as 52 given that 36 (half of 72) are expected is negligible
($P=0.0\%$).

\item Comparing the space densities of the \emph{mvCG}s and the \emph{vHCG}s,
   we deduce that the HCG catalogue is only 8\% complete.
A comparison of the parameter distributions between the \emph{mvCG}s and
\emph{vHCG}s indicates that the HCG is incomplete in groups of small 
angular sizes, high fraction of light in the brightest galaxy,
as well as (for the C06 model) faint brightest galaxy
magnitudes and low surface brightness.

\item We find that the velocity filtering of \emph{mpCG}s
does not necessarily imply that the resulting accordant-velocity CGs are 
physically dense. 
We tested different criteria to classify the accordant-velocity CGs according 
to the maximum 3D galaxy separations and the line of sight elongations.   
We find that, with the most conservative criterion, \emph{at least 3/5 of the 
mock accordant-velocity CGs are physically dense}, 
although the precise fraction depends on the galaxy 
formation model used. 

\item The large majority of non-Real mock accordant-velocity 
CGs are caused by chance alignments within 
larger groups, rather than within larger regions such as large-scale filaments.
 
\item We find that the fraction of chance alignments decreases from 28--47\% for
the particle-based \emph{pmvCG}s to 24--41\% for the sample with close pairs
removed (\emph{mvCG}s) to only 14--30\% once we fold in the biases of the HCG
(\emph{mvHCG}s). 
This explains in part why simulation studies \citep{Mamon86,WM89,HKW95} predict
more chance alignments than one infers from actual observations.

\end{enumerate}

Table~\ref{samples} indicates that typically half of the groups are lost once we
apply the criterion to blend close projected pairs of galaxies. While the discarded
groups no longer satisfy the selection threshold of 4 galaxies within 3 magnitudes from the
brightest, there are still many blended pairs within the groups that survived
the blending criterion (in 1/3, 1/4, and 1/8 of the mock CGs for B06, DLB,
and C06, respectively,
with
similar fractions for \emph{mpCG}s and \emph{mvCG}s). Could this mean that a
significant fraction of
the \emph{pHCG}s and \emph{vHCG}s contain blended galaxies?

A comparison of the lists of \cite{Hickson82} (selected from photographic plates)
and \cite{HKA89} (CCD-based) indicates that the latter found 10 HCGs with extra galaxies (24, 26
[+3], 27, 43, 51, 70, 72, 76[+2], 83 and 99), plus one with one galaxy less
(HCG 40). 
So 10\% of the original HCG groups contained blended galaxies as discovered
with the better CCD photometry. It therefore appears that our fractions of
1/8 to 1/3 of \emph{mpCG}s with blended pairs is high. 
But if the truth is in between our particle case and our extended case, then
we should have of order 6--17\% of groups with blended galaxies, which is in
rough agreement with what we see in the HCGs.

Now, three HCGs have been observed at much higher resolution with HST imaging.
One shows no extra galaxies (HCG 87), while the other two show more
interacting units than counted by \cite{HKA89}: HCG 31 
seems to have 7 galaxies and not just 4, while HCG 90 has 5 galaxies and not
4. Binomial statistics suggest that, with 95\%
confidence, the view of 2 groups out of 3 with extra galaxies implies that the
fraction of such groups with blended pairs is between 14\% and 86\%. But HST
is probably biased towards dense interacting HCGs. Still, there could be
interacting pairs showing galaxies that have been blended even with the CCD
images of \cite{HKA89}. So, with the high resolution of the HST, 
the fraction of \emph{pHCG}s and \emph{vHCG}s
with blended pairs may be considerably higher than 10\% and in agreement with
the fractions found in the mock CGs.

We can also estimate the number of HCGs that are
physically dense groups of at least 4 galaxies.
As discussed in Sect.~\ref{hcgsample}, the HCG catalogue 
has 100 members, among which 99 are compact groups 
(since HCG~54 is a collection of \hii regions), 
of which only 83 actually fulfil the original
magnitude concordance criterion ($R$-band magnitude range less than 3).
Among the 72 \emph{pHCG}s whose brightest
and faintest galaxies are brighter than $R=14.44$ and $R=17.44$,
respectively, only 52 (72\%) have at least 4 accordant velocities, and among these,
we expect roughly between 36 and 44 
HCGs that are physically dense groups of at least 4 galaxies.
Extrapolating to the 68 accordant-velocity HCGs (including those with
magnitude range greater than 3 mags), 
\emph{we expect no more than $\sim$ 58 physically dense HCGs with at least 4
galaxies}.  

In
comparison, \cite{Mamon86} had predicted that 47 out of what he thought would
be 78 accordant velocity HCGs are caused
by chance alignments (60\%), while the remaining 40\% 
are physically dense (but he predicted that
half of these dense groups were 
unbound systems). 
We are therefore less pessimistic than 
\cite{Mamon86}
on the fraction of chance alignments polluting the HCG
catalogue, since chance alignments appear to represent between 14\% and 30\%  
of the \emph{mvHCG}s (we were not able to check what fraction of the Real ones are
unbound: see Appendix~\ref{appbind}).
Part of this discrepancy is caused by Mamon's (1986) reliance on
simulations without consideration of selection effects such as observers
blending close projected pairs of galaxies. 
Still, the percentage of chance alignments in the particle mock
velocity-filtered compact groups (\emph{pmvCG}s) is only 28--47\% (depending
on the SAM). Nevertheless, 
given the wide range of chance alignments fractions among the
three SAMs, 
one cannot
rule out that a more realistic galaxy formation model would lead to as much
as 60\% of chance alignments.

\subsection{Comparison with McConnachie et al.}
\label{compare}

\cite{McCEP08} (MEP) have 
published a study very similar to
ours: they also extracted \emph{pmpCG}s from the DLB model obtained from the
Millennium dark matter simulations. Their sample extended to $r = 18$, which
corresponds to roughly one-quarter of a magnitude fainter than our limit of 
$R < 17.44$. Their other \emph{pmpCG} criteria appear to be
almost exactly the same as ours (following the criteria of \citealp{Hickson82}), 
although their algorithm works differently \citep{McCPES09}. 
MEP found a total of over 15\,000 \emph{pmpCG}s over $4\pi\,\rm sr$. 

Using precisely the same input galaxy catalogue (from  \citealp{Blaizot+05}) as
MEP, we find 25\,000 \emph{pmpCG}s, so, our algorithm is nearly 1.6 times more
efficient than MEP's in finding \emph{pmpCG}s. 
Surprisingly, if we build a light cone as we did in Sect.~\ref{testing}, 
with apparent magnitude limit of $R=17.67$ ($\simeq r_{\rm SDSS}=18$), 
we obtain a mock galaxy catalogue that is 3 times denser than the
\citeauthor{Blaizot+05} mock galaxy  catalogue used by MEP. 
From our mock galaxy catalogue, we extract 15\,191 \emph{pmpCG}s in 1.2693 sr, 
which means that, with the data used in this work, we are
$\sim 10$ times more efficient than MEP in finding CGs, principally by
  differences in the parent samples of galaxies, but also by differences 
in the CG detection algorithm.

MEP and us agree that a significant fraction of \emph{mpCG}s are caused by chance
alignments: MEP found 71\% of their \emph{pmpCG}s are CAs while we find 80\%
(with our hybrid classification, see Table~\ref{nattab}).

There are, however, several important differences between our two studies:
\begin{itemize}
\item MEP build their sample from a mock that extends beyond the box size of
  the Millennium simulation --- using the output of the Mock Map Facility
  (MoMaF) code of \cite{Blaizot+05}, while our mock galaxy catalogues are
  limited to the size of the simulation box. 
  However, as shown in Sect.~\ref{testing}, 
  working on a light cone or working on a single simulation box leads to
  similar 
numbers of mock CGs (we identify a factor of 1.2 fewer \emph{mpCG}s, and 1.5
more \emph{mvCG}s). 
\item MEP consider CGs with a faint magnitude limit, while we also tie in a
  bright magnitude limit to ensure that all mock CGs were built from galaxies
  that spanned a range of over 3 magnitudes.
\item We have analysed the galaxies from 3 different SAMs, while MEP have
  only considered the DLB sample, which we found to produce the smallest
  fraction of physically dense \emph{mpCG}s.
\item MEP only provide statistics for the mock CGs defined in projection
  (\emph{pmpCG}s, which they refer to as `HA's) but 
do not consider the subset of accordant-velocity groups (\emph{pmvCG}s). 
We think this would have been worthwhile
because ever since \cite{HMdOHP92} published the HCG galaxy redshifts,
most analyses have thrown out the discordant velocity HCGs. 
\item MEP did not consider selection effects, while we considered both the
  galaxy confusion from close, blended, projected pairs, as well as the
  biases that we determined for the Hickson's visual selection of the HCGs.

\item MEP only considered those \emph{mpCG}s with $k\geq 3$  galaxies that
  lie very close in real space,
  while we considered 
$k \ge 4$ (to be consistent with Hickson's initial
  motivation to have at least 4 galaxies per HCG).
\item MEP define the Real \emph{mpCG}s using a Friends-of-Friends linking
  length in real space, while we use a maximum real-space
  separation and the elongation along the 
line-of-sight. Structures built from small numbers of components 
with Friends-of-Friends algorithms tend to be more filamentary
(e.g. \citealp*{MFW93}).  
For \emph{mpCG}s that are CALGs
  or CAFs, the most distant outlier will determine a similar maximum length and
  critical linking length. However, for \emph{mpCG}s without outliers
  (e.g. Real \emph{mpCG}s), the linking length will be smaller than the
  3D length. In other words, selecting Real groups with a linking length of
  $200 \, h^{-1} \, \rm kpc$ will result in group 3D lengths considerably
  greater.
Moreover, for \emph{mpCG}s with both
foreground and background galaxies, MEP's $\ell$ must be compared to 
our \emph{half}-maximum size $s_{\rm cut}/2$, and there is here a discrepancy
of a factor two.
Worse, for those (admittedly rare) cases of, say 4, 
galaxies aligned
along the line of sight at roughly equal separations just below $\ell$, one
will end up with a group that spans up to $3\,\ell = 600 \, h^{-1} \, \rm
kpc$, which is now three times our maximum 3D length, but will still be 
called \emph{Compact Association} (Real) by MEP, although it clearly is a chance
alignment. 
In summary, this point and the previous one imply that 
MEP's criterion for calling an \emph{mpCG} Real is much more
liberal than ours.

\end{itemize}

\subsection{Perspectives}
In forthcoming papers, we will analyse the distribution of and correlations between the physical
characteristics of the \emph{mvCG}s, and show
how they depend on their classification, in view of optimising the
probability that a CG selected in redshift-space is physically dense.
It would be worthwhile to probe the formation of the physically dense CGs  
by analysing the 
merger trees of galaxies in the mock CGs.
Finally, the analysis presented here will need to be confirmed with increasingly realistic
simulations of galaxy catalogues, for example constructed from future galaxy
formation models run on the recent high resolution Millennium-II dark matter
simulation \citep{BoylanKolchin+09}, and also on future
high-resolution cosmological 
hydrodynamical simulations, with realistic prescriptions for feedback from
AGN and supernovae. 

\section{Acknowledgements}
We thank Paul Hickson, Abilio Mateus and Jack Sulentic for useful exchanges
and an anonymous referee for interesting comments.
The Millennium Simulation databases used in this paper and the web
application providing online access to them were constructed as part of the
activities of the German Astrophysical Virtual Observatory.
We thank Richard Bower, Darren Croton, and Gabriele De Lucia for allowing
public access for the
outputs of their very impressive
semi-analytical models of galaxy formation.
This research has made use of VizieR database maintained by the Center de
Donn\'ees Stellaires in Strasbourg, France, and the NASA/IPAC Extragalactic Database (NED)
which is operated by the Jet Propulsion Laboratory, California Institute of
Technology, under contract with the National Aeronautics and Space
Administration.
This work was partially supported by the European Commission's ALFA-II program
through its funding of the Latin-American European Network for
Astrophysics and Cosmology (LENAC), Consejo de Investigaciones Cient\'{\i}ficas y
T\'ecnicas de la Rep\'ublica Argentina (CONICET), the Agencia
Nacional de Promoci\'on Cient\'{\i}fica and the Secretar\'{\i}a de
Ciencia y T\'ecnica de la Universidad Nacional de C\'ordoba (SeCyT).

\bibliography{cgs}
\appendix
\section{From r-SDSS to R-Johnson apparent magnitude limit}
\label{appmags}
According to Table~3 of 
\cite*{FSI95}, $r$--$R$ has typical values of
0.36 (E), 0.31 (S0), 0.33 (Sab), 0.32 (Sbc), 0.30 (Scd), and
0.20 (Im), and we adopt $\langle r-R \rangle = 0.33$. We thus restrict the HCG
sample to a total extrapolated $R$ band extinction-corrected 
magnitude 
\begin{eqnarray}
R_T^0 &\!\!=\!\!& B_T^0-(B\!-\!R)_{\rm iso} + \left (B_T\!-\!B_T^0\right )\,\left (1-{A_R/A_V\over
  A_B/A_V}\right ) \label{RT0} \\
&\!\!<\!\!& 17.44 \ , \nonumber
\end{eqnarray}
where we adopted $A_B/A_V=1.33$ and $A_R/A_V=0.75$ \citep{CCM89}.
Equation~(\ref{RT0}) assumes no $B$--$R$ colour gradient in the galaxy.

\section{Group binding energies}
\label{appbind}
The
binding energy of the particles of a mock CG is difficult to determine
because it is not clear which particles belong to the CG.

One could alternatively use the galaxies instead of the particles. But most
of the mass of a group is thought to lie in between the galaxies, so one must
be careful on how the binding energy analysis is performed.
We separate the group into the system of galaxies and the remaining
intergalactic dark matter.
The kinetic energy of the group would then be
\begin{equation}
K = K_{\rm g} + K_{\rm d} = {3\over 2}\,\left (M_{\rm g} + M_{\rm d}\right)\,\sigma_v^2 \ ,
\label{K}
\end{equation}
where $M_{\rm g}$ and $M_{\rm d}$ are the masses of the galaxies and of the intergalactic
dark matter,
respectively, while $\sigma_v$ is the one-dimensional velocity dispersion
assumed to be the same for the intergalactic dark matter and the galaxies, and where we
placed ourselves in the group centre-of-mass to get rid of the bulk kinetic energy.
The potential energy is more difficult to handle as it is the sum of the
potential energies of the galaxy system, the intergalactic dark matter system and the
cross-term between galaxies and the intergalactic dark matter. Suppose a group has a factor
$\mu$ times more intergalactic dark matter than mass in galaxies. From
equation~(\ref{K}), the kinetic
energy is then 
\begin{equation}
K=(\mu+1)\,K_{\rm g} \ .
\label{K2}
\end{equation}
If the galaxies and the dark matter have
similar distributions in space (which is not really possible, since by
definition the intergalactic dark matter is outside the galaxies), then the
total potential energy can be written
\begin{eqnarray}
W &=& W_{\rm g,g} + W_{\rm d,d} + W_{\rm g,d} \nonumber \\
&=& - {G M_{\rm g}^2 \over r_{\rm G}} - {G M_{\rm d}^2 \over r_{\rm G}} - {G M_{\rm g} M_{\rm d} \over r_{\rm G}} \nonumber \\
&=& \left(\mu^2+\mu+1\right)\,W_{\rm g,g} \ ,
\label{W}
\end{eqnarray}
where $r_{\rm G}$ is the gravitational radius, assumed to be the same for the three
terms.
So, if the group is in virial equilibrium, one has $2\,K + W = 0$, then
according to equations~(\ref{K2}) and (\ref{W}), one finds that
\begin{equation}
2\,(\mu+1)\,K_{\rm g} + \left (\mu^2+\mu+1\right)\,W_{\rm g,g} = 0
\end{equation}
Therefore, one finds that 
\begin{equation}
-{2 K_{\rm g,g}\over W_{\rm g,g}} = {\mu^2+\mu+1\over\mu+1}
\label{virrat} \ ,
\end{equation}
so, if the intergalactic dark matter makes up for say $\mu=4$ times as much
as the galaxy mass, then if the group is in virial equilibrium,
equation~(\ref{virrat}) leads to $-2\,K_{\rm g,g}/W_{\rm g,g} = 21/5 = 4.2$. 
Therefore, \emph{the virial ratio of the galaxies in a group can be far off
  from unity!}
It will depend on the
fraction of mass in galaxies (i.e. on $1/(\mu+1)$, hence on $\mu$). Finally,
since the virial ratio of the galaxy system appears to be greater than 2, we
would incorrectly conclude that typical groups are unbound!

One way to avoid these problems is to assign to each galaxy the fraction of the
total group mass equal to the ratio of its mass divided by the total mass in
galaxies for that group. In other words, we are putting the intergalactic
dark matter mass in each galaxy in proportion to its mass.
This is equivalent to $\mu=0$, hence to a virial ratio of unity according to
equation~(\ref{virrat}). 

The simulation data that we have at our disposal provides the virial masses
of the PG3D groups (dark matter included). If a mock CG is a CA, then we cannot
know how much dark matter is assigned to this CA, but only to the PG3D group
associated with it. We therefore choose to scale the galaxy masses to the
PG3D group mass, i.e.
\begin{equation}
m' = {m\over \sum_{i\in{\rm PG3D}} m_i}\,M_{\rm PG3D} \ . 
\label{mcorr}
\end{equation}
Still, there remains the issue of CG galaxies that do not belong to any PG3D.
One possibility is to apply equation~(\ref{mcorr}) to the CG galaxies that
lie within PG3D groups, without scaling the masses of the isolated galaxies.
The alternative is to scale by the fraction of mass in the whole
simulation box, instead of the PG3D group.

Another issue is that the virial spheres around the galaxies will tend to
overlap inside the Real CGs. One therefore needs to soften the galaxy-galaxy
potential energy of interaction, for example with the approximation \citep{Mamon87}:
\begin{equation}
V_{\rm int}\left(r_{i,j}\right) = - {G m'_i m'_j \over \sqrt{r_{i,j}^2+r_{i,j,{\rm rms}}^2}}
\end{equation}
where $r_{i,j,{\rm rms}}$ is the root mean squared of the half-mass radii of the
pair of galaxies $\{i,j\}$.
The virial theorem becomes
\begin{equation}
2\,K + \sum {\bf F} \cdot {\bf r} = 0 \ ,
\end{equation}
where the Clausius virial of a group is
\begin{eqnarray}
\sum {\bf F} \cdot {\bf r} &=& \sum_i {\bf r}_i \cdot \nabla V_{\rm
  int}\left(r_{i,j}\right) \nonumber \\
&=& \sum_i r_i \cdot \sum_{j\neq i} {\bf r}_{i,j} {G m'_i m'_j \over \left
  (r_{i,j}^2+r_{ij,{\rm rms}}^2\right)^{3/2}} \nonumber \\
&=& G \sum_{i<j} m'_i m'_j
  {r_{i,j}^2 \over \left(r_{i,j}^2+r_{i,j,{\rm rms}}^2\right)^{3/2}}
\end{eqnarray}
where ${\bf r}_{i,j}$ is the vector separating
galaxies $i$ and $j$.

But how do we estimate the galaxy half-mass radii? We can compute
analytically the half-mass radius of the matter within the virial radius, say
for an NFW model, with a concentration $c=r_{\rm v}/r_{-2}=10$ (where
$r_{-2}$ is the `scale' radius of slope $-2$), for which $r_{\rm h}/r_{\rm v} \simeq 0.36$
(\citealp{LM01}, Fig.~4 and eq.~[28]).  But we could also compute the
half-mass radius within a larger radius, say the turnaround radius beyond
which the Universe is expanding, and which is typically 3.5 times the virial
radius. Assuming that the NFW model extends that far (see
\citealp{Prada+06}), going to the turnaround radius amounts to increasing the
concentration by a factor of $r_{\rm ta}/r_{\rm v} \approx 3.5$. So, if
$c=10$ for a galaxy, at the turnaround radius, we would use $c=35$ and find
$r_{\rm h}/r_{\rm ta} \simeq 0.23$, i.e. $r_{\rm h}/r_{\rm v} \simeq 0.79$.

But then, how do we estimate the mass within the virial radius of the galaxy?
We could guess a mass-to-light ratio $M(r_{\rm v})/L_B = 100$, although 
$M(r_{\rm v})/L_B$ is thought to decrease
with increasing luminosity to reach a minimum around 70 \citep{Eke+06}. 
Then $r_{\rm v} =
\left[2/\Delta \left(GM/H_0^2\right)\right]^{1/3} = \left[2/\Delta \left(G
  (M/L)L/H_0^2\right)\right]^{1/3} = 544 
\,(L/10^{11})^{1/3}\,\rm kpc$ for $\Delta=100$ (as we used in the paper),
$H_0 = 73 \,\rm km \,s^{-1} \, Mpc^{-1}$ 
(as
used in the Millennium Simulation) and $M/L=100$. For $L_*=0.18 \times
10^{11}\,L_\odot$, 
we end up
with $r_{\rm v}=307\,\rm kpc$. So for $L_*$ galaxies, we need a softening of
typically $r_0 = 70$ to 250 kpc, i.e. $c_0 = r_0/L^{1/3} = 0.03$ to 0.10. 
Of course, the higher the softening scale, the less
negative is the potential energy and the less bound is the system.

We test our prescription by computing the virial ratios of the PG3Ds using
both PG3D scaling and box scaling of the galaxy masses, with different
values for the softening scale $c_0$. The correct scaling must lead to virial
ratios of unity, independent of group mass.

\begin{figure}
\centering
\includegraphics[width=\hsize]{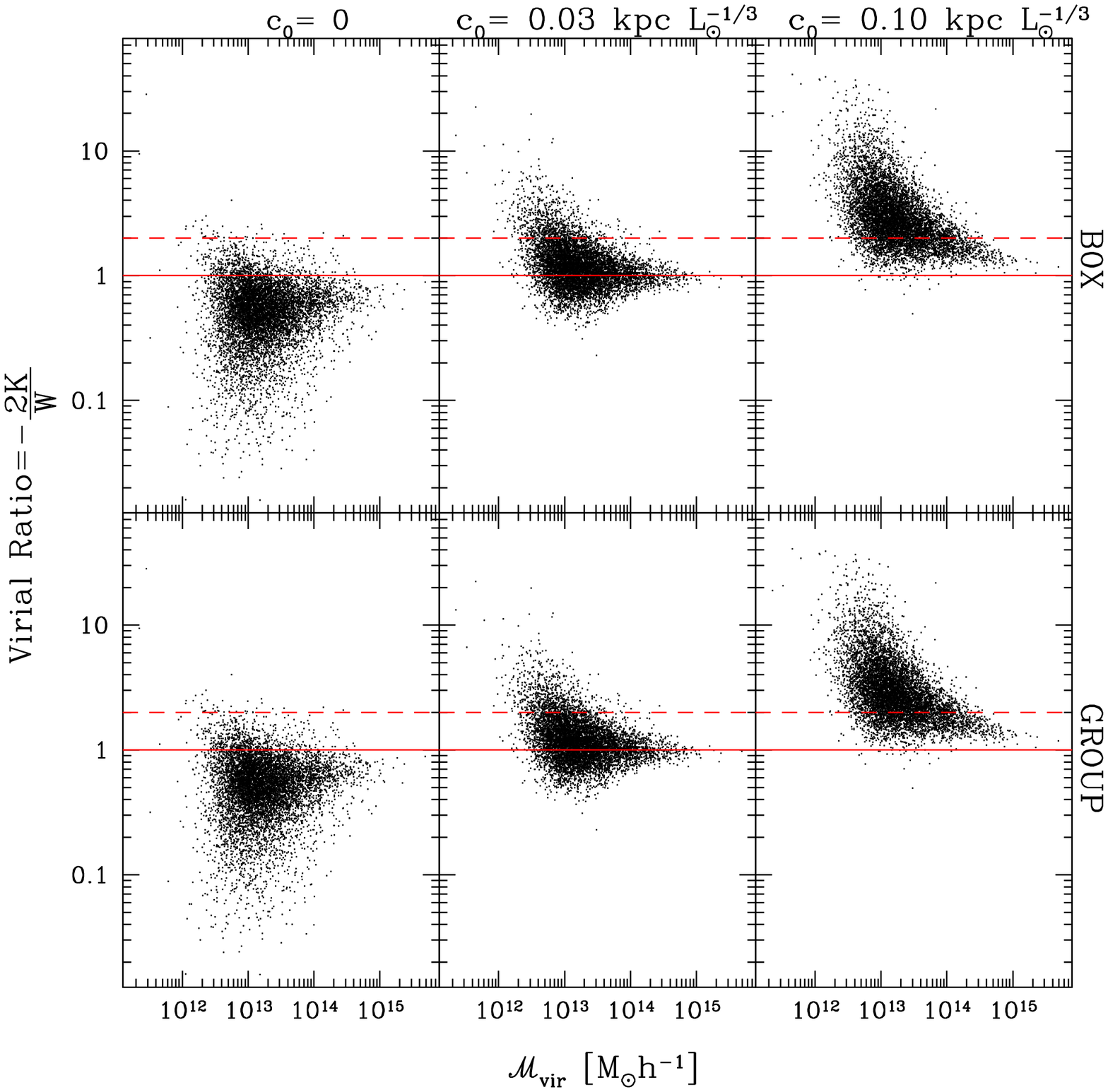}
\caption{Virial ratio as a function of virial mass for PG3D groups with
  $N>10$ members extracted from DLB model, where the galaxy mass
correction (eq.~[\ref{mcorr}]) is over the entire box (\emph{top panel}) or
the parent group (\emph{bottom panel}).
The \emph{solid red line} shows the expected virial ratio of unity, while 
the \emph{dashed line} shows the limit for unbound groups.
\label{virial_ratio}}
\end{figure}

Figure~\ref{virial_ratio} shows the results of our test on PG3D groups.
For both normalisations, we find that the softening $c_0 = 0.03\,{\rm
  kpc}\,L_\odot^{-1/3}$ (corresponding to $r_0 = 70 \, h^{-1} \, \rm kpc$ for
$L=L_*$ galaxies) bring the virial ratios of the highest mass PG3Ds to
unity. Without the softening, the potential energies are overestimated (in
absolute value), hence
the virial ratios are underestimated, while with too strong softening the
virial ratios are overestimated.

However, Figure~\ref{virial_ratio} indicates that even with the correct
softening, i.e. with correct virial ratios at the high mass end, there is so
much scatter in the virial ratios at low masses, that over 13\% of the PG3Ds
are found to be unbound for PG3D virial masses below
$4\times10^{13}\,M_\odot$. 
This means that for the typical masses of the CGs, our virial ratio
estimator is too inaccurate to use as a CG classifier.

\end{document}